\documentclass[a4paper,11pt]{article} 

\usepackage{epsfig}
\usepackage{amsmath}
\usepackage{cite}
\usepackage{axodraw}
\usepackage{lscape}
\usepackage{multirow}
\usepackage{graphics,amssymb,epsfig,float,amstext, epstopdf}
\usepackage{graphicx}
\usepackage{pdflscape}
\usepackage{color}
\usepackage{jheppub}
\allowdisplaybreaks

\usepackage{graphicx}
\usepackage{comment}
\usepackage{bm,array}
\usepackage{spverbatim}
\usepackage{listings}
\usepackage{graphics}
\usepackage{epsf}
\usepackage{color}
\usepackage{amsmath}
\usepackage{amssymb}
\usepackage{latexsym}
\usepackage{slashed}
\usepackage{float}
\usepackage{cases}
\usepackage{multirow}
\def\ba{\begin{array}}
\def\ea{\end{array}}

\usepackage{listings}
\usepackage{color}

\definecolor{mygreen}{rgb}{0,0.6,0}
\definecolor{mygray}{rgb}{0.5,0.5,0.5}
\definecolor{mymauve}{rgb}{0.58,0,0.82}

\definecolor{mygreen}{rgb}{0,0.6,0}
\definecolor{mygray}{rgb}{0.5,0.5,0.5}
\definecolor{mymauve}{rgb}{0.58,0,0.82}

\lstnewenvironment{CPP}
{
\lstset{ %
	backgroundcolor=\color{white},   
	basicstyle=\footnotesize\ttfamily,        
	breakatwhitespace=false,         
	breaklines=true,                 
	captionpos=b,                    
	commentstyle=\color{mygreen},    
	deletekeywords={...},            
	escapeinside={\%*}{*)},          
	extendedchars=true,              
	frame=none,	                   
	keepspaces=true,                 
	columns=flexible,
	keywordstyle=\color{blue},       
	language=C++,                 
	morekeywords={*,...},           
	numbers=left,                    
	numbersep=5pt,                   
	numberstyle=\tiny\color{mygray}, 
	rulecolor=\color{black},         
	showspaces=false,                
	showstringspaces=false,          
	showtabs=false,                  
	stepnumber=0,                    
	stringstyle=\color{mymauve},     
	tabsize=2,	                   
	title=\lstname                   
}
}
{}
\lstnewenvironment{F77}
{
\lstset{ %
	backgroundcolor=\color{white},   
	basicstyle=\footnotesize\ttfamily,        
	breakatwhitespace=false,         
	breaklines=true,                 
	captionpos=b,                    
	commentstyle=\color{mygreen},    
	deletekeywords={...},            
	escapeinside={\%*}{*)},          
	extendedchars=true,              
	frame=none,	                   
	keepspaces=true,                 
	columns=flexible,
	keywordstyle=\color{blue},       
	language={[77]Fortran},            
	morekeywords={*,...},           
	numbers=left,                    
	numbersep=5pt,                   
	numberstyle=\tiny\color{mygray}, 
	rulecolor=\color{black},         
	showspaces=false,                
	showstringspaces=false,          
	showtabs=false,                  
	stepnumber=0,                    
	stringstyle=\color{mymauve},     
	tabsize=2,	                   
	title=\lstname                   
}
}{}
\def\tp{t^{\prime}}
\def\sp{s^{\prime}}
\def\alphas{\alpha_s}
\def\rinv{R^{-1}}

\def\kkgl{g^\star_1}
\def\qp{q^\prime}

\def\td{t^{\prime}_D}
\def\ts{t^{\prime}_S}
\def\qbar{\bar{q}}
\def\ubar{\bar{u}}
\def\dbar{\bar{d}}
\def\qbarp{\bar{q}^\prime}

\def\qkd{q^\bullet_{1}}
\def\ukd{u^\bullet_{1}}
\def\dkd{d^\bullet_{1}}
\def\qks{q^\circ_{1}}
\def\uks{u^\circ_{1}}
\def\dks{d^\circ_{1}}
\def\qkb{q^*_{1}}

\def\ubkd{\bar{u}^\bullet_{1}}
\def\dbkd{\bar{d}^\bullet_{1}}

\def\dbks{\bar{d}^\circ_{1}}
\def\qbkb{\bar{q}^*_{1}}

\def\qkps{q^{\prime \circ}_{1}}

\def\qkpb{q^{\prime *}_{1}}

\def\qbkps{\bar{q}^{\prime \circ}_{1}}

\def\qbkpb{\bar{q}^{\prime *}_{1}}
\def\mnkk{m_{K}}
\def\amgl{m_{g^\star_1}}
\def\amqk{m_{q^*_1}}

\def\amqd{m_{q^\bullet_1}}
\def\amqs{m_{q^\circ_1}}

\def\tp{t^\prime}
\def\up{u^\prime}
\def\rinv{R{^{-1}}}
\def\thetaw{\theta_W}
\def\alphaem{\alpha_{em}}
\def\alphau1{\alpha_1}
\def\alphasu2{\alpha_2}
\def\alphas{\alpha_s}
\newcommand{\beq}{\begin{equation}}
\newcommand{\eeq}{\end{equation}}
\newcommand{\bea}{\begin{eqnarray}}
\newcommand{\eea}{\end{eqnarray}}
\newcommand{\mpt}{{/\!\!\!\! \vec{P}_T}} 
\newcommand{\met}{{\slashed{E}_T}} 

\title{LHC Collider Phenomenology of Minimal Universal Extra Dimensions}

\author[a,b,c]{Jyotiranjan Beuria,}
\author[a,b]{AseshKrishna Datta,}
\author[d]{Dipsikha Debnath,}
\author[d]{Konstantin T.~Matchev}

\affiliation[a]{Harish-Chandra Research Institute, Allahabad 211019, India}
\affiliation[b]{Homi Bhabha National Institute, Training School Complex, Anushakti Nagar, \\
                Mumbai 400085, India}
\affiliation[c]{Regional Centre for Accelerator-based Particle Physics, Harish-Chandra Research Institute, \\ Allahabad 211019, India}
\affiliation[d]{Physics Department, University of Florida, Gainesville, FL 32611, USA}

\preprint{HRI-P-17-02-001 \\
\vspace*{-0.8cm}
\begin{flushright}
RECAPP-HRI-2017-002
\end{flushright}
}

\abstract{We discuss the collider phenomenology of the model of Minimal Universal Extra Dimensions (MUED) at the Large hadron Collider (LHC).
We derive analytical results for all relevant strong pair-production processes of two level 1 Kaluza-Klein partners and use them to
validate and correct the existing MUED implementation in the fortran version of the {\sc Pythia} event generator. 
We also develop a new implementation of the model in the C++ version of {\sc Pythia}.  
We use our implementations in conjunction with the {\sc Checkmate} package to derive the LHC bounds on MUED 
from a large number of published experimental analyses from Run 1 at the LHC.
}

\begin{document} 
\maketitle
\flushbottom

\section{Introduction}
\label{sec:intro}

The ongoing run of the Large Hadron Collider (LHC) at CERN is continuing its quest for new physics Beyond the Standard Model (BSM).
From a theorist's viewpoint, the first step in this program is to characterize the anticipated new physics in terms of its particle content, interactions, and 
parameter values at the relevant energy scale of the collider experiment \cite{Ask:2012sm}.\footnote{Two alternative model-independent 
approaches, where one does {\em not} have to specify a particular BSM model, have recently been proposed in 
Refs.~\cite{Debnath:2014eaa,Debnath:2015wra}.}  Traditionally, this can be done in one of two ways: by considering a ``complete" model, 
or a ``simplified" model.
\begin{itemize}
\item {\em Complete models.} Here one introduces the complete new physics structure which exists below the ultraviolet cutoff of the 
effective theory. The standard and most popular example of this type is the framework of supersymmetry (SUSY) \cite{Martin:1997ns}. 
Unfortunately, this approach usually brings about a large number of input parameters; for example, even the minimal version of SUSY, 
the Minimal Supersymmetric Standard Model (MSSM), has more than 100 input parameters \cite{Haber:1993wf}, which need 
to be trimmed down to a more manageable number. This can be done in two ways:
\begin{enumerate}
\item {\em Phenomenological models.} Here, one retains full generality, but fixes some subset of the original input parameters to 
some typical values which are motivated or dictated by the current experimental limits. 
For example, the general MSSM parameter space is severely constrained by
the existing bounds from CP-violating and flavor-changing processes, thus a reasonable and conservative approach is to set the 
corresponding CP-violating and flavor-changing parameters to zero, achieving a drastic reduction in the dimensionality of the 
relevant parameter space \cite{Djouadi:1998di,Berger:2008cq}. This simplification has allowed the LHC collaborations to publish 
several analyses from SUSY searches in the thus obtained ``phenomenological MSSM" (pMSSM) 
\cite{Aad:2015baa,Khachatryan:2016nvf,Aaboud:2016wna}.
\item {\em Models with high-scale boundary conditions.} Alternatively, one can choose to specify the model parameters 
at the matching scale, where the ultraviolet completion of the theory is expected to impose certain restrictions on the allowed pattern of
parameters. The most celebrated example of this type is a grand unification theory (GUT), which unifies the values for the gauge 
and certain Yukawa couplings at the GUT scale. In SUSY GUTs, the soft masses for the SUSY partners are also unified according to
irreducible representations of the GUT group, as in the minimal supergravity (mSUGRA) model, 
for example \cite{Inoue:1982pi,Chamseddine:1982jx,Barbieri:1982eh,Hall:1983iz}.
However, this approach also requires a procedure to relate the handful of theory input parameters at the high scale to the 
phenomenological model parameters at the electroweak  scale. The procedure involves numerically solving the
renormalization group equations, minimizing the effective potential, and computing the radiative corrections to 
the particle mass spectrum. In the case of SUSY, several programs have been developed to perform these tasks:
{\sc ISAJET} \cite{Baer:1999sp}, {\sc SOFTSUSY} \cite{Allanach:2001kg}, {\sc SuSpect} \cite{Djouadi:2002ze}, and {\sc SPheno} \cite{Porod:2003um}. 
Since the calculations are done at a fixed order in perturbation theory, some differences between the 
programs inevitably arise, due to a different set of higher order terms being neglected. 
In the past, these differences were carefully scrutinized for the case of the MSSM, 
which led to an estimate of the inherent theoretical systematic error in 
these types of calculations in SUSY \cite{Allanach:2003jw,Belanger:2005jk}.
\end{enumerate}
\item {\em Simplified models.} Given that phenomenological models have too many input parameters, while
models with high-scale boundary conditions involve unverified theoretical assumptions, the recent trend has been to
consider the so-called simplified models, where one focuses only on one specific event topology of interest,
and treats the masses of the BSM particles in it (as well as the overall rate) as free parameters \cite{Alves:2011wf}.
This approach is a compromise between the previous two --- on the one hand, the number of input parameters is greatly 
reduced, since only a few BSM states are present in a given event topology, while at the same time, the mass parameters 
can be treated in full generality, without any a priori theoretical assumptions. This has motivated the 
LHC collaborations to publish simplified model interpretations of their new physics searches as well
\cite{ATLAS:2011ffa,Okawa:2011xg,Chatrchyan:2013sza,CMS:2016swi}.
The main goal of the simplified model approach is to allow for published LHC results 
to be easily recast for a variety of BSM scenarios, e.g.~with different couplings, spins, etc. 
However, all these benefits do come at a certain cost --- for any given complete model, the limits derived from 
an exclusive analysis of a single simplified model topology will be very conservative, since 
each subprocess typically contributes only a small fraction to the rate for the inclusive final state signature being studied.
The bounds are therefore generally underestimated and to obtain the true limit in a given complete model, one has to 
combine the results from several simplified model analyses \cite{Gutschow:2012pw,Barnard:2014joa}, reintroducing the
problems mentioned above.
\end{itemize}

For many years, the only complete models which were seriously investigated by the Tevatron and LHC collaborations, were 
different versions of low energy supersymmetry: mSUGRA, gauge-mediated SUSY \cite{Giudice:1998bp}, 
anomaly-mediated SUSY \cite{Randall:1998uk,Giudice:1998xp}, etc. Apart from its attractive features from a theorist's point of view,
the ubiquity of supersymmetry was also partially due to the fact that it was readily available in the existing Monte Carlo simulation 
tools used by the experimental collaborations. However, with the recent progress on automation and standardization\footnote{For example, 
the original SUSY Les Houches accord \cite{Skands:2003cj,Allanach:2008qq} has now been extended to general BSM models 
\cite{Alwall:2007mw,Degrande:2011ua,Brooijmans:2012yi}.} of the
simulation chain leading from a theory Lagrangian to fully simulated events \cite{Ask:2012sm}, this restriction has largely been lifted,
and a much wider class of models can now be explored. 

Our main objective in this paper is to consider a complete model which is different from supersymmetry, and study its LHC phenomenology. 
We choose the scenario of Universal Extra Dimensions (UED) \cite{Appelquist:2000nn} in its minimal version, 
Minimal UED (MUED) \cite{Cheng:2002iz}. Minimal UED bears some similarities with supersymmetry: for example, each particle 
of the Standard Model has heavier Kaluza-Klein (KK) partners with the same gauge quantum numbers. 
Yet there are important differences as well: e.g., the spins of the KK partners differ by $1/2$ from those of the superpartners.
Unfortunately, this feature is not unique to UED, but is also shared, e.g., by Little Higgs models with $T$-parity \cite{Cheng:2003ju}.
Furthermore, the spin determination of new particles in cascade decays is a notoriously difficult task
\cite{Athanasiou:2006ef,Wang:2006hk,Burns:2008cp}, which is expected to take some time after the initial discovery. 
Thus, given the experimental challenges in discriminating SUSY from UED \cite{Battaglia:2005zf,Smillie:2005ar,Datta:2005zs,Datta:2005vx}, 
we believe that MUED should be seriously considered as a ``complete model" benchmark, on equal footing with SUSY.

At this point in time, studies of MUED are additionally motivated by the null results from the new physics searches at the LHC.
In the last couple of years, there has been a flurry of activity in designing models which ``hide" the new physics from the LHC.
One of the standard methods for doing so is to arrange for a ``compressed" mass spectrum with a mass degeneracy 
of the relevant SUSY particles, so that the resulting decay products are too soft to be triggered upon and tagged 
in the experimental analysis \cite{Martin:2007gf,Baer:2007uz,Schwaller:2013baa,Han:2014kaa,Han:2014aea,An:2016nlb,Konar:2016ata}, or a ``stealth" mass spectrum, 
where the new physics signature becomes identical to the SM background,
since the additional particles are too soft to make any appreciable difference
\cite{Fan:2011yu,Alves:2012ft,Macaluso:2015wja,Cheng:2016npb}. 
In models with many input parameters like the MSSM, such degeneracies, while phenomenologically 
desirable, from a theoretical point of view appear accidental and unmotivated.
In contrast, at each KK level, the new particle mass spectrum in MUED is very degenerate, 
and for a good reason --- the dominant contribution to the mass comes from the compactification,
$m\sim R^{-1}$, where $R$ is the size of the extra dimension.
Therefore, MUED represents a well motivated benchmark for a complete model with mass degeneracies.

In this paper, we shall pursue several different goals:
\begin{enumerate}
\item {\em Reliable MUED event generation.} By now MUED is available in several parton level event generators.
The original implementation \cite{Datta:2010us} was done directly in {\sc CalcHEP} \cite{Pukhov:2004ca} 
and {\sc CompHEP} \cite{Pukhov:1999gg}, while the more modern implementations in
{\sc Madgraph} \cite{Alwall:2011uj} and {\sc Whizard} \cite{Kilian:2007gr,Christensen:2010wz} have interfaced to the 
{\sc FeynRules} package \cite{Alloul:2013bka}. Since parton level event generators exclusively 
define the final state at the parton level, they are ideally suited for simplified model studies, while 
complete models are traditionally more conveniently studied with general purpose event generators, 
which eventually have to be used as part of the simulation chain anyway. While MUED is available in
{\sc Herwig++} \cite{Bahr:2007ni} and {\sc Sherpa} \cite{Hoche:2014kca}, here we shall focus on the 
case of {\sc Pythia} given its wide use in the experimental community. The original implementation \cite{azuelos,Allanach:2006fy,ElKacimi:2009zj}
was done in the fortran version {\sc Pythia6} \cite{Sjostrand:2006za} which is now being phased out.
One of our major goals here is to create a modern implementation of MUED in {\sc Pythia8} \cite{Sjostrand:2007gs}, 
as well as to validate and fix\footnote{The history of the MUED implementation in {\sc Pythia6} is somewhat complicated.
The first papers to calculate the matrix elements \cite{Macesanu:2002db,azuelos} contained some typos, 
some of which were eventually fixed in an erratum several years later \cite{Nandi2}. Unfortunately,
by then the work on implementing UED in {\sc Pythia} had already started \cite{Allanach:2006fy} 
and those corrections did not propagate to the official version \cite{ElKacimi:2009zj}.} 
for backward compatibility the current {\sc Pythia6} version, which has a couple of issues. 
First, the matrix elements incorporated in {\sc Pythia} were computed in the degenerate limit, ignoring
the mass splittings among the KK partners at a given KK level. Here we will account for 
the correct mass splittings among the level 1 KK partners. The corresponding analytical results are  
given in Appendix~\ref{app:matsq-ndg}, while in Section~\ref{sec:degenerate} we also quote
the corresponding expressions for the degenerate mass limit, correcting some previous results in the literature. 
Second, we shall be careful to include all strong production subprocesses. Finally, 
we shall expand the nomenclature to allow separate treatment of KK partners with 
different gauge quantum numbers. The details of our {\sc Pythia6} ({\sc Pythia8}) implementation 
can be found in Appendix~\ref{sec:pythia6} (Appendix~\ref{sec:pythia8}).
\item {\em Estimate the systematic uncertainties in the UED mass spectrum calculation.} Even though
the formulas for the one-loop corrections to the KK mass spectrum have been known for a while
\cite{Cheng:2002iz}, when implementing them in an event generator, one has to face several decisions which 
may affect the results. In the spirit of refs.~\cite{Allanach:2003jw,Belanger:2005jk}, in Section~\ref{sec:mass} 
we shall compare the MUED mass spectra resulting from the existing implementations 
in {\sc Pythia} and {\sc CalcHEP} and identify the origin of the observed differences.
\item {\em Estimate the relative importance of level 2 KK resonances for the inclusive production of level 1 KK particles.} 
While our main focus will be on strong pair-production of level 1 KK partners, in Section~\ref{sec:level2} we shall also 
evaluate the relative contributions from diagrams with level 2 KK partners. 
\item {\em Apply {\sc Checkmate} to the case of MUED.} At the moment, there exist several packages
which allow testing of BSM models against published LHC results, e.g. 
{\sc Checkmate} \cite{Drees:2013wra}, {\sc SModelS} \cite{Kraml:2013mwa},
{\sc Fastlim} \cite{Papucci:2014rja}, {\sc Madanalysis5} \cite{Conte:2014zja}, etc.
However, most of the previous applications in the literature have been limited to some version of supersymmetry.
Here we would like to showcase a different complete model, namely MUED, using the 
latest version of {\sc Checkmate} \cite{Dercks:2016npn}.
\item {\em Derive the current LHC bounds on minimal UED.} Qualitatively, the collider signatures of
MUED are similar to those of supersymmetry with conserved $R$-parity and a neutral LSP
\cite{Cheng:2002ab}. For example, for every simplified model event topology in SUSY, 
there is an analogous one in MUED. However, there are important quantitative differences as well.
First, because of the relatively small mass splittings among the KK partners, the visible 
SM decay products tend to be rather soft, which would hurt the efficiency. This may be offset by the fact that 
the KK quarks are fermions, and thus have larger cross-sections compared to the spin 0 squarks in SUSY.
Finally, the mass degeneracy of the KK partners implies that there are many relevant subprocesses,
and therefore, a large number of final state signatures which one could potentially target. An important
theoretical question, therefore, is what is the optimal search strategy for MUED in terms of
the discovery signature. A tool like {\sc Checkmate} is perfect to answer this question.
\end{enumerate}

Previously there have been several related studies of MUED phenomenology at hadron colliders (see \cite{Macesanu:2005jx,Servant:2014lqa} 
for reviews). In general, the main characteristic feature of the MUED collider signature depends on whether the 
lightest KK particle (LKP) is allowed to decay to SM particles through some sort of KK-parity violating interactions 
or not \cite{Rizzo:2001sd,Macesanu:2002ew}. Here we shall focus on the most challenging scenario in which the 
LKP is stable and neutral, as in supersymmetry, and consider pair-production of level 1 KK partners\footnote{Unlike the case of SUSY,
in UED one could also look for the {\em single} production of level 2 KK resonances 
\cite{Datta:2005zs,Chang:2011vt,Kim:2011tq,Chang:2012wp,Edelhauser:2013lia,Datta:2013lja,Shaw:2014gba,Dey:2014ana}, 
which is beyond the scope of this paper.}. 
Then the generic signal is missing transverse energy, accompanied by 
(typically soft) decay products from KK cascade decays.
Given the softness of the signatures, the original papers focused at first on the clean leptonic channels, 
which represent golden channels for discovery \cite{Cheng:2002ab,Lin:2005ix,Kazana:2007zz,Ribeiro:2009hsa,Bhattacharyya:2009br,Belyaev:2012ai}.
Subsequently, multilepton signatures with jets were also considered 
\cite{Bhattacherjee:2009jh,Choudhury:2009kz,Bhattacherjee:2010vm}, and
even hadronic multijet channels \cite{Murayama:2011hj,Datta:2011vg,Cacciapaglia:2013wha} were studied. 
In the extreme case of very small mass splittings, where none of the decay products can be reliably reconstructed, 
it has been suggested that one could usefully use a jet from initial state radiation \cite{Choudhury:2016tff}
or analyze the pattern of reconstructed tracks \cite{Chakraborty:2016qim}.
In Section~\ref{sec:bounds} we shall use the full suite of LHC analyses
implemented in {\sc Checkmate} to derive LHC limits on the MUED scale $R^{-1}$ from Run 1 data,
which can be compared to published ATLAS results \cite{TheATLAScollaboration:2013uha,Aad:2015mia,Yamaguchi:2012uva,Morvaj:2014xsh}. 
Such LHC results have interesting cosmological implications, since they test the idea that Kaluza-Klein dark matter 
is a viable component of the dark matter in the Universe \cite{Arrenberg:2008wy}. 
Section~\ref{sec:conclusions} is reserved for our conclusions.

\section{MUED preliminaries}

\subsection{Notations and conventions}

To the extent that we are interested in LHC phenomenology, our focus here will be on the strongly interacting 
KK partners: the KK gluons, $g^\star_n$, and the KK quarks, $q^\bullet_n$ and $q^\circ_n$, where the index 
$n$ labels the KK level. We follow the notation of \cite{Macesanu:2002db}, where the KK excitation of an  
$SU(2)$-doublet fermion of the SM is denoted with a bullet (a filled circle), while the KK excitation of an
$SU(2)$-singlet fermion of the SM is denoted with an open circle (see Table~\ref{table:spectra}). If the
$SU(2)$ quantum number of the KK quark could be either one, we shall use an asterisk as a placeholder.   

\begin{table}[t]
\centering
{\small
\begin{tabular}{|c|c|c|c|}
\hline
  Level 1  particle     &   Notation       & \multicolumn{2}{c|}{Mass  (GeV)}    \\
\cline{3-4}
  &   &       {\sc CalcHEP}       &   {\sc Pythia6} \\
  \hline
\hline 
 KK gluon          &      $g_1^\star$    &  640.49      &       622.55     \\   
\hline
\hline
KK doublet quark  ($u$- or $d$-type) &      $q_1^\bullet$  &  --      &       --          \\
\hline
KK doublet $u$ quark  &   $u_1^\bullet$   &   597.57     &    584.76     \\
\hline
KK doublet $d$ quark   &  $d_1^\bullet$   &    597.57      &     584.76    \\
\hline
\hline
KK singlet quark ($u$- or $d$-type)  & $q_1^\circ$      &   --      &    --      \\
\hline
KK singlet $u$ quark   & $u_1^\circ$      &  586.24      &    574.23        \\
\hline
KK singlet $d$ quark    &   $d_1^\circ$    & 584.42       &       572.53          \\
\hline
\hline
Generic KK quark & $q_1^\ast$  & --     &       --     \\
\hline
\end{tabular}
}
\caption{\label{table:spectra} 
Notation for the colored level 1 KK partners and their masses obtained with default settings
from {\sc CalcHEP} and {\sc Pythia} for $R^{-1}=500$ GeV and $\Lambda R=20$.} 
\end{table}

While the SM quarks are chiral and receive their tree-level masses from the Higgs mechanism, 
the KK modes $q^\bullet_n$ and $q^\circ_n$ (with $n\ge 1$) have both a left-handed and a right-handed component,
and obtain their masses predominantly from the compactification:
\beq
m_{q^\ast_n}=\sqrt{(nR^{-1})^2+m^2_{q_0}}, \quad {\rm for}\ n\ge 1.
\eeq
Since the KK modes at level 1 (i.e.,~with $n=1$) are the lightest, they are most easily produced at the LHC,
and so will be the main focus of our discussion in the next Section.
In MUED, KK parity is conserved, and level 1 gluons $g^\star_1$ and level 1 quarks $q^\ast_1$ are 
necessarily pair-produced. They will then cascade decay to the LKP, which in MUED is the level 1 KK mode 
$B_1$ of the SM hypercharge gauge boson.

\subsection{Mass spectrum uncertainties}
\label{sec:mass}

Obtaining an accurate particle mass spectrum in a given new physics
scenario is the first step in any robust phenomenological study. 
In MUED, all boundary terms are assumed to vanish at some matching scale $\Lambda > R^{-1}$,
but get regenerated at lower scales $\mu$ from RGE running, which leads to a set of
``boundary" mass corrections to the level 1 KK modes \cite{Cheng:2002iz}
\bea
\bar\delta m_{q^\bullet_n} &=&
\frac{n}{R}
\left(
3\frac{g^2_3}{16\pi^2}+\frac{27}{16}\frac{g^2_2}{16\pi^2}+\frac{1}{16}\frac{g^{\prime 2}}{16\pi^2}
\right)
\ln \frac{\Lambda^2}{\mu^2},  \label{bardmq}\\
\bar\delta m_{u^\circ_n} &=&
\frac{n}{R}
\left(
3\frac{g^2_3}{16\pi^2}+\frac{g^{\prime 2}}{16\pi^2}
\right)
\ln \frac{\Lambda^2}{\mu^2},  \label{bardmu}\\
\bar\delta m_{d^\circ_n} &=&
\frac{n}{R}
\left(
3\frac{g^2_3}{16\pi^2}+\frac{1}{4}\frac{g^{\prime 2}}{16\pi^2}
\right)
\ln \frac{\Lambda^2}{\mu^2},  \label{bardmd}
\\
\bar\delta m_{g^\star_n}^2 &=&
\frac{n^2}{R^2}
\frac{23}{2} \frac{g^2_3}{16\pi^2}
\ln \frac{\Lambda^2}{\mu^2}.  \label{bardmg}
\eea
In addition, the KK gluons also receive corrections from bulk loops \cite{Cheng:2002iz}
\beq
\delta m_{g^\star_n}^2 =
-\frac{3}{2} \frac{\zeta(3) g^2_3}{16\pi^4}
\frac{1}{R^{-2}},  \label{dmg}
\eeq
which are not logarithmically enhanced and are generally numerically quite small.

In the last two columns of Table~\ref{table:spectra} we show the mass spectrum for the
level 1 colored KK partners which results from the application of the mass corrections (\ref{bardmq}-\ref{dmg})
in the two packages {\sc CalcHEP} and {\sc Pythia6}, respectively. In both cases, we use default settings and 
input the same MUED parameters: $R^{-1}=500$ GeV and $\Lambda R = 20$, 
yet the calculated masses are noticeably different.
In particular, the mass spectrum obtained in {\sc Pythia6} appears to be slightly lighter.
The origin of this difference can be tracked back to the way in which the two programs compute 
the values for the SM gauge couplings ($g_3,g_2,g'$) which enter eqs.~(\ref{bardmq}-\ref{dmg})
and the renormalization scale $\mu$ at which (\ref{bardmq}-\ref{dmg}) are applied.
\begin{itemize}
\item In {\sc Pythia6}, the MUED spectrum is computed in the {\tt PYUEDC} subroutine. 
The running strong coupling $g_3(Q)$ is evaluated at the 
scale $Q=\rinv$ through the {\sc Pythia} function {\tt PYALPS}, using the default $\Lambda_{QCD}$ (=250 MeV) 
and a lowest order (LO) running. 
However, the electroweak couplings are not evolved, and {\tt PYUEDC} 
uses a fixed input $\alphaem =1/137$ through the default setting of the parameter 
{\tt PARU(101)}. The values of $g_2$ and $g'$ which are needed in (\ref{bardmq}-\ref{bardmd})
are then derived from the fixed inputs for $\alphaem$ and the Weinberg angle, $\thetaw$. 
\item
In contrast, the MUED formulation for {\sc CalcHEP} runs either all three couplings (with the
parameter {\tt RG} set to 1) or none (with {\tt RG}$=$0, which is the default setting). 
For the case of running couplings, the scale $\mu$ at which the couplings are evaluated is set by the 
parameter {\tt scaleN}: with {\tt scaleN}$=$1, it is set at $\mu=\rinv$, while
{\tt scaleN}$=$2 (the default) implies $\mu=2\rinv$. When {\tt RG}$=$0, the three gauge couplings 
$\alpha_i$ are fixed at their values at $\mu=M_Z$ through the input variables 
{\tt c1MZ}, {\tt c2MZ} and {\tt c3MZ} in CalcHEP \cite{Datta:2010us}.
\end{itemize}
It is clear that by default, not all three gauge couplings are consistently run to the same renormalization scale
in the two programs. In principle, the differences are higher order, and are expected 
to be of the same size as the neglected next-to-leading order corrections in (\ref{bardmq}-\ref{dmg}). 
As a sanity check, we have confirmed that when the gauge couplings are treated the same way, i.e.
\begin{itemize}
\item we let all three gauge couplings run at the same time, 
\item we use the same starting values at $M_Z$,
\item the running is done at leading order,
\item the couplings are evolved up to the same renormalization scale $\mu=\rinv$,
\end{itemize}
the two programs give identical results for the masses of the level 1 KK particles. 
This exercise required appropriate tweaking of the code and/or some of the program parameters. 
Note that the choice of renormalization scale $\mu$ used for the computation of
the KK mass spectrum may in principle be different from the choice of $\mu$ 
used for the gauge couplings appearing in the scattering (decay) vertices. Those couplings need to be 
evaluated at a scale compatible with the set-up for the hard-scattering, particularly 
since the parton distribution in use provides the value of $\alphas$ compatible with
its parametrization.

In summary, the differences seen in Table~\ref{table:spectra} between the default mass spectrum calculations in {\sc Pythia} and {\sc CalcHEP}
are due to the effect from higher order terms, and are thus indicative of the theoretical uncertainty 
of these calculations. In what follows, we shall sometimes need to compare cross-section results 
from {\sc Pythia} and {\sc CalcHEP} for the same KK spectrum. To do so, for concreteness we shall
first obtain the default {\sc Pythia} mass spectrum and then feed those masses to {\sc Calchep},
rather than the other way around --- this is easier operationally, since in {\sc CalcHEP} the UED masses 
can be easily set externally, without hacking the internal code.

\section{Strong production in MUED}

In this section we discuss strong production in pure QCD of level 1 KK partners and validate its implementation in {\sc CalcHEP} and {\sc Pythia}.
At parton level, these processes were first computed in \cite{Macesanu:2002db} in the degenerate limit
\beq
m_{g^\star_1} = m_{q^\ast_1} \equiv m_{K},
\label{degenerate}
\eeq
where one neglects the mass differences among the different level 1 KK quarks and/or the KK gluon.
The calculation was repeated in Ref.~\cite{azuelos}, whose analytical results were then used for the 
MUED implementation in {\sc Pythia6} \cite{Allanach:2006fy,ElKacimi:2009zj}.
Since the mass splittings among the level 1 KK partners arise from renormalization effects, (\ref{degenerate})
is not a bad approximation, but we would like to avoid it nevertheless.  Thus we have revisited the calculation 
of the parton-level pair production cross-sections for various pairs of colored level 1 KK partners in the general case of arbitrary 
mass splittings. The corresponding results are listed in Appendix~\ref{app:matsq-ndg}. The formulas are rather long and not very instructive,
but we shall make use of them later on when we create our own MUED implementation in {\sc Pythia6}, as explained in Appendix \ref{sec:pythia6}.
Here (in Section \ref{sec:degenerate}) we quote the simplified version of our results from Appendix~\ref{app:matsq-ndg} in the degenerate
limit (\ref{degenerate}), which can be compared with the previous literature. This will also allow us to test the current version of {\sc Pythia6}
in Section~\ref{sec:validation}. Finally, in Section~\ref{sec:level2} we close this section with a comment on the relative importance of diagrams 
mediated by level 2 KK particles. 

\subsection{Analytical results in the limit of degenerate level 1 KK partners}
\label{sec:degenerate}

In this subsection we list the simplified version of our results from Appendix~\ref{app:matsq-ndg} in the degenerate limit (\ref{degenerate}).
The formulas are written in terms of the Mandelstam variables $s$, $t$ and $u$. 
In the degenerate case (\ref{degenerate}), the two final state particles have the same mass $m_K$, 
and the Mandelstam variables obey the identity
\beq
s+ t+ u = 2m_K^2.
\label{stu}
\eeq
This is why it is convenient to redefine
\bea
t' &\equiv& t-m_K^2, \\ [1mm]
u'&\equiv& u-m_K^2,
\eea
after which eq.~(\ref{stu}) simplifies to
\beq
s+ t'+ u' = 0.
\label{stpup}
\eeq

The spin-averaged squared matrix elements for the different $2\to2$ strong production processes are given by
\begin{eqnarray}
\overline{\sum} |{\cal M}|^2_{g g \to g_1^\star g_1^\star}
  & = &
\frac{9 \alphas^2}{16 (s \tp \up)^2} \
\left(s^2 + {\tp}^2 + {\up}^2 \right) \nonumber \\ 
 && \qquad  \quad \times \
\left[6 \mnkk^4 s^2 - 6\mnkk^2 s \tp\up + 3{\tp}^2 {\up}^2 \
 + 2 s^2 \left({\tp}^2 + {\up}^2 \right)\right],  \label{ggg1g1} \\ [2mm]
\overline{\sum} |{\cal M}|^2_{q g \to \qkb g_1^\star} 
  &= & 
\frac{-\alphas^2}{72 s \tp {\up}^2} \
\left(2s^2 + 2{\tp}^2 + {\up}^2\right) \left(9s^2 + 9{\tp}^2 - {\up}^2\right),  \label{qgq1g1}  \\ [2mm]
\overline{\sum} |{\cal M}|^2_{q q \to \qkb \qkb} 
  &=&
\frac{\alphas^2}{54 {\tp}^2 {\up}^2}\
\left[{\tp}^4 + {\up}^4 - s^4 + 6 s^2 ({\tp}^2 + {\up}^2) \
 + \mnkk^2 (6 {\tp}^3 + 6 {\up}^3 -s \tp \up ) \right], ~~~~  \label{qqq1q1} \\ [2mm]
\overline{\sum} |{\cal M}|^2_{q \qp \to \qkb \qkpb} 
  &=&
\frac{\alphas^2}{18{\tp}^2} \left(-4 \mnkk^2 s+4 s^2+{\tp}^2\right),  \label{qqpq1qp1} \\ [2mm]
\overline{\sum} |{\cal M}|^2_{g g \to \qkb \qbkb} 
  &=& 
\frac{-\alphas^2}{24 s^2 {\tp}^2 {\up}^2} \
\left(4 s^2 - 9 \tp \up \right) \left[ 4 \mnkk^4 s^2 - 4 \mnkk^2 s \tp \up - \up \tp \left({\tp}^2 + {\up}^2 \right)\right],  \label{ggq1q1bar} \\ [2mm]
\overline{\sum} |{\cal M}|^2_{q \qbar \to \qkb \qbkb} 
  &=&
\frac{\alphas^2}{54 s^2 {\tp}^2}
\left[12 \mnkk^2 s \left(s^2-s \tp+4 {\tp}^2\right) \right. \nonumber\\
 && \qquad \qquad \left. +12 s^4+16 s^3 \tp+23 s^2 {\tp}^2 \
+36 s {\tp}^3+48 {\tp}^4\right],   \label{qqbarq1q1bar} \\ [2mm]
\overline{\sum} |{\cal M}|^2_{q \qbarp \to \qkd \qbkps} 
  &=&
\frac{\alphas^2}{18 {\tp}^2} \left(-4 \mnkk^2 s+4 s^2+{\tp}^2\right),  \label{qqpbarq1qp1bards}  \\ [2mm]
\overline{\sum} |{\cal M}|^2_{q \qbarp \to \qkb \qbkpb} 
  &=&
\frac{\alphas^2}{18 {\tp}^2} \left(4 \mnkk^2 s+4 s^2+8 s \tp+5 {\tp}^2\right),  \label{qqpbarq1qp1bar} \\ [2mm]
\overline{\sum} |{\cal M}|^2_{q q \to \qkd \qks} 
  &=& 
\frac{\alphas^2}{9 {\tp}^2 {\up}^2} \ 
\left[2 \mnkk^2 s \left({\tp}^2 + {\up}^2 \right) + 2 s^4 \
 - 8 s^2 \tp \up + 5 {\tp}^2 {\up}^2\right],   \label{qqq1q1ds} \\ [2mm]
\overline{\sum} |{\cal M}|^2_{q \qp \to \qkd \qkps} 
  &=&
\frac{\alphas^2}{18 {\tp}^2} \left(4 \mnkk^2 s+4 s^2+8 s \tp+5 {\tp}^2\right), \label{qqpq1qp1ds} \\ [2mm]
\overline{\sum} |{\cal M}|^2_{q \qbar \to \qkpb \qbkpb} 
  &=&
\frac{4 \alphas^2}{9 s^2} \left(2 \mnkk^2 s+ {\tp}^2 + {\up}^2 \right), \label{qqbarqp1qp1bar}\\ [2mm]
\overline{\sum} |{\cal M}|^2_{q \qbar \to \kkgl \kkgl} 
  &=&
\frac{-\alphas^2}{27 s^2 {\tp}^2 {\up}^2} 
\left(4 s^2-9 \tp \up \right) 
\left[ \left(5 s^2-12 \tp \up \right) 
\left(\mnkk^2 s-\tp \up \right)-4 \mnkk^2 s^3\right].
\label{qqbarg1g1}
\end{eqnarray}

Several sanity checks of these formulas are possible. Each expression has a prefactor whose denominator indicates the type of 
relevant diagrams. For example, one can verify that processes 
(\ref{ggg1g1}), (\ref{qgq1g1}), (\ref{ggq1q1bar}) and (\ref{qqbarg1g1}) are mediated by $s$-, $t$- and $u$-channel diagrams;
processes (\ref{qqq1q1}) and (\ref{qqq1q1ds}) have $t$- and $u$-channel contributions;
process (\ref{qqbarq1q1bar}) has  $s$- and $t$-channel diagrams;
processes (\ref{qqpq1qp1}), (\ref{qqpbarq1qp1bards}), (\ref{qqpbarq1qp1bar}) and (\ref{qqpq1qp1ds}) are mediated by a single $t$-channel diagram,
while (\ref{qqbarqp1qp1bar}) is a pure $s$-channel process.
Second, processes with identical initial or final state particles should be invariant under $u\leftrightarrow t$.
It is easy to see that the respective processes 
(\ref{ggg1g1}), (\ref{qqq1q1}), (\ref{ggq1q1bar}), (\ref{qqq1q1ds}) and (\ref{qqbarg1g1})
indeed obey this condition.

\subsection{{\sc Pythia6} validation}
\label{sec:validation}

The parton-level results (\ref{ggg1g1}-\ref{qqbarqp1qp1bar}) from the previous subsection can be used to test
the {\sc Pythia6} implementation\footnote{The process of eq.~(\ref{qqbarg1g1}), $q \qbar \to \kkgl \kkgl$, was not included in {\sc Pythia6}
since at the LHC it was expected to have a relatively small cross-section in comparison to $gg\to \kkgl\kkgl$ of eq.~(\ref{ggg1g1}).} 
of MUED \cite{Allanach:2006fy,ElKacimi:2009zj} 
which was done in the same degenerate limit (\ref{degenerate}).
To be more precise, in {\sc Pythia6} the approximation (\ref{degenerate}) is applied only when computing the event weight
in the {\tt PYXUED} subroutine, which introduces a common mass variable {\tt xmnkk} for all level 1 KK particles
(the same as the common mass $m_K$ used in eqs.~(\ref{ggg1g1}-\ref{qqbarg1g1}) above).
On the other hand, the treatment of the phase-space weights in the Pythia subroutine
{\tt PYSIGH} correctly accounts for the actual masses of the individual KK excitations
as calculated in the {\tt PYUEDC} subroutine. Thus, there appears to be an inconsistency in the way 
these two basic parts of the {\sc Pythia6} code operate, and our goal in this subsection will be to 
quantify the effect and ultimately rectify this inconsistency. Along the way, we shall also validate the 
programmed matrix elements in {\sc Pythia6} and perform the necessary corrections.

\begin{table}[t]
\centering
{\small
\begin{tabular}{||c|c||c||c||c|c||c|c||}
\hline \hline
\multicolumn{2}{||c||}{Parton-level}  & Mass &  {\sc CalcHEP}  &  \multicolumn{4}{|c||}{{\sc Pythia6}}   \\
\cline{5-8} 
\multicolumn{2}{||c||}{subprocess}  & parameters &   & default  & diff   & modified & diff \\
\cline{1-2}
ISUB  & name &        &  $\sigma$ (fb) & $\sigma$ (fb) & $\%$ &  $\sigma$ (fb) &  $\%$ \\ [2mm]
\hline
\hline
%
311 & $gg \to g_1^\star g_1^\star$ & $m_{g_1^\star}$ & 783.38  & 810.34 & 3.44 & 787.7 & 0.55\\[2mm]
\hline
312 & ($qg \to \qkb g_1^\star$) & $m_{q_1^\ast}$, $m_{g_1^\star}$  &      &      &      &          &          \\
    & $ug \to \ukd g_1^\star$  & & 1040.3 & 1893.36 & 82.0 & 1057 & 1.59 \\
\hline
313a &  ($qq \to \qkb \qkb$) & $m_{q_1^\ast}$, $m_{g_1^\star}$  &      &     &     &     &     \\
     &  $uu \to \ukd \ukd$    & & 549.6 & 1258.23 & 128.94 & 752.3 & 31.14 \\
313b & ($q \qp \to \qkb \qkpb$) & $m_{q_1^\ast}$, $m_{q_1^{\prime \ast}}$, $m_{g_1^\star}$&      &     &     &     &     \\
     & $u d \to \ukd \dkd$    &  & 288.8  & 551.70 & 91.03 & 295.2 & 2.19 \\
\hline
314 & ($gg \to \qkb \qbkb$)& $m_{q_1^\ast}$  &      &      &      &      &      \\
    & $gg \to \ukd \ubkd$  &  & 14.78 & 6.7 & 54.67 & 13.88 & 6.28 \\
\hline
315 & ($q\qbar \to \qkb \qbkb$)& $m_{q_1^\ast}$,  $m_{g_1^\star}$ &      &      &       &      &      \\
    & $u\ubar \to \ukd \ubkd$ &  & 47.81 & 82.14 &  71.80 & 58.5 & 20.11 \\
\hline
316 & ($q\qbarp \to \qkd \qbkps$)& $m_{q_1^\bullet}$,  $m_{q_1^{\prime \circ}}$, $m_{g_1^\star}$&      &      &      &      &      \\
    & $u \dbar \to \ukd \dbks$  &   & 52.03 & 85.13 & 63.62 & 53.25 & 2.31 \\
\hline
317 & ($q\qbarp \to q_1^\ast \qbkpb$) & $m_{q_1^\ast}$,  $m_{q_1^{\prime \ast}}$, $m_{g_1^\star}$ &      &      &      &      &      \\
    & $u\dbar \to \ukd \dbkd$     &  &   34.92   & 57.97 & 66.01 & 36.37 & 4.07 \\
\hline
318a & ($qq \to \qkd \qks$) &  $m_{q_1^\bullet}$, $m_{q_1^\circ}$, $m_{g_1^\star}$ &      &      &      &      &      \\
     & $uu \to \ukd \uks$  &  & 1027.8 & 1474.94 & 43.50 & 1010 & 1.73 \\
318b & ($q \qp \to \qkd \qkps$) & $m_{q_1^\bullet}$, $m_{q_1^{\prime \circ}}$, $m_{g_1^\star}$    &      &       &       &       &       \\
     & $u d \to \ukd \dks$   &   & 211.6 &  353.12     & 66.88    & 218.5   & 3.2 \\
\hline
319 & ($q \qbar \to \qkpb \qbkpb$)& $m_{q_1^{\prime \ast}}$ &      &      &      &      &      \\
    & $u \ubar \to \dkd \dbkd$ &    & 13.87 & 12.52 & 9.73 & 13.85 & 0.14 \\
%
\hline
\hline
\end{tabular}
}
\caption{\label{table:pythia} Comparison of the MUED cross sections for different subprocesses
at the 7 TeV LHC, obtained with two different MUED implementations: {\sc CalcHEP} \cite{Datta:2010us} and 
{\sc Pythia6} \cite{Allanach:2006fy,ElKacimi:2009zj}. In both programs, we choose the {\tt CTEQ5L} parton distribution functions with $Q^2=\hat{s}$
and use the default mass spectrum as calculated in {\sc Pythia} with $R^{-1}=500$ GeV, $\Lambda R=20$. 
The first two columns identify the parton level subprocess, while
the third column lists the mass parameters which enter the corresponding full matrix element (see Appendix~\ref{app:matsq-ndg}).
The fourth (fifth) column gives the default result from {\sc CalcHEP} ({\sc Pythia6}), and the sixth column gives 
the percent difference between them. The last two columns give the corresponding results from a similar comparison with a 
modified version of {\sc Pythia6} with a suitable choice of the parameter ${\tt xmnkk}$ as explained in the text.}
\end{table}

In Table \ref{table:pythia} we compare 7 TeV LHC cross-section results from the default MUED implementations 
in {\sc CalcHEP} \cite{Datta:2010us} (fourth column) and in {\sc Pythia6} \cite{Allanach:2006fy,ElKacimi:2009zj} (fifth column)
on a process by process basis. In each case, we use the same KK mass spectrum as calculated in {\sc Pythia6} for 
$R^{-1}=500$ GeV, $\Lambda R=20$. The first two columns identify the parton level subprocess, which in {\sc Pythia6} 
is labelled by the {\tt ISUB} variable.
The second column of Table \ref{table:pythia} lists the generic subprocess. For ease of comparison, 
we do not sum over quark flavors, but choose a specific flavor selection as indicated in the table.
In general, the full matrix element for each subprocess (see Appendix~\ref{app:matsq-ndg}) 
carries dependence on the masses of one or more KK particles, as indicated in the third column of Table \ref{table:pythia}.
In both {\sc Pythia6} and {\sc CalcHEP} we use {\tt CTEQ5L} parton distribution functions and use the
value of $\alpha_s$ returned by setting the renormalization scale $Q^2=\hat{s}$, which is the default in {\sc CalcHEP}.

The sixth column of Table \ref{table:pythia} displays the percent difference of the results.
We notice that the default answers from {\sc CalcHEP} and {\sc Pythia6} can be quite different, 
which prompts an investigation into the potential origin of this mismatch. Since we have matched 
the mass spectra by hand, we have already ruled out the mass uncertainties discussed previously 
in Section~\ref{sec:mass}. 

The main culprit for the discrepancy is the use of the degenerate mass approximation (\ref{degenerate}).
By design, the MUED matrix elements in {\sc Pythia} were programmed in the degenerate mass
limit (\ref{degenerate}), which was adopted in Refs.~\cite{Macesanu:2002db,azuelos}. In contrast, 
the matrix elements in {\sc CalcHEP} are computed with the correct mass spectrum, including 
the proper 1-loop mass splittings. We also note that the common KK mass {\tt xmnkk} 
in {\sc Pythia} by default is set equal to the mass of the lightest KK particle, 
the level 1 KK mode of the hypercharge gauge boson, $B_1$. This value is then used to compute
the cross-sections for all subprocesses, including strong production.
In reality, the scale for the strong production cross-sections is set by the masses of the {\em colored} 
level 1 KK partners which are heavier than $B_1$ by $10-20\%$ \cite{Cheng:2002iz}. 
This would lead to a systematic overestimation of the strong cross-sections in {\sc Pythia}, 
which is indeed what we observe\footnote{There are two exceptions to this trend: the process $q \qbar \to \qkpb \qbkpb$
with ${\tt ISUB}=319$, whose cross-section scales as $1/s$ instead of $1/m_K^2$, see (\ref{qqbarqp1qp1bar}), and
the process $gg \to \qkb \qbkb$ with ${\tt ISUB}=314$, for which the {\sc Pythia6} code contains a typo, as we will discuss below.} 
in Table \ref{table:pythia}.

Having identified the problem, the question now is whether one can still recover the correct answer in {\sc Pythia}
without reprogramming the code with the full matrix elements from Appendix~\ref{app:matsq-ndg},
and instead simply making a judicious choice of the value of the common mass {\tt xmnkk}. 
This seems to be plausible in the case of subprocesses where the full matrix element 
depends on a single mass parameter, namely the mass of the (identical) final state particles. 
Table \ref{table:pythia} contains three such processes, ${\tt ISUB}=\{311,314,319 \}$.

Let us illustrate the procedure with the example of KK gluon pair production where the
matrix element depends on a single parameter, $m_{g_1^\star}$. If we substitute $m_K=m_{g_1^\star}$
in the degenerate case formula (\ref{ggg1g1}), we obtain
\bea
\overline{\sum} |{\cal M}|^2_{g g \to g^\star g^\star}
  & = &
\frac{9 \alphas^2}{16 (s \tp \up)^2} \
(s^2 + {\tp}^2 + {\up}^2) \nonumber \\ 
 && \qquad  \times 
\left[6 m_{g_1^\star}^4 s^2 - 6m_{g_1^\star}^2 s \tp\up + 3{\tp}^2 {\up}^2 \
 + 2 s^2 ({\tp}^2 + {\up}^2)\right],
\eea 
where now $\tp=t-m_{g_1^\star}^2$ and $\up=u-m_{g_1^\star}^2$. 
The result is identical to eq.~(\ref{Mnd311}), which holds in the non-degenerate limit.
This demonstrates that for single-parameter matrix elements, we should be able to recover 
the correct answer with the proper choice of the parameter {\tt xmnkk} in {\sc Pythia}. 
The last two columns in Table \ref{table:pythia} show that this trick generally works: 
after modifying the default choice for {\tt xmnkk}, we get excellent agreement for two 
out of the three processes, namely ${\tt ISUB}=311$ and ${\tt ISUB}=319$. 
On the other hand, the discrepancy for ${\tt ISUB}=314$ is still sizable, on the order of $6\%$,
and deserves further scrutiny. The subprocess is $g g \to {\qkb} {\qbkb}$, thus we need to replace 
$m_K\to m_{q_1^\ast}$ everywhere in (\ref{ggq1q1bar}), obtaining
\beq
\overline{\sum} |{\cal M}|^2_{g g \to {\qkb} {\qbkb}}  = 
\frac{-\alphas^2}{24 s^2 {\tp}^2 {\up}^2} 
\left(4 s^2 - 9 {\tp} {\up}\right) \left[4 m_{q_1^\ast}^4 s^2 - 4 m_{q_1^\ast}^2 s {\tp} {\up} - {\up} {\tp} \left({\tp}^2 + {\up}^2\right)\right],
\eeq
which is the same answer as (\ref{Mnd314}), and the simple fix above should have been sufficient.
Due to the identical final state particles, the expression (\ref{Mnd314}) is symmetric with respect to $u\leftrightarrow t$, but
as it turns out, the {\sc Pythia6} code is not. This indicates a problem, which was apparently inherited from the 
expressions in Ref.~\cite{azuelos}.  

For processes whose matrix elements depend on 2 or 3 mass parameters (see the third column in Table~\ref{table:pythia}), 
a valid procedure for lifting the degenerate limit expressions (\ref{ggg1g1}-\ref{qqbarg1g1})
to their non-degenerate counterparts from Appendix~\ref{app:matsq-ndg} is not available. 
Short of implementing the full expressions from Appendix~\ref{app:matsq-ndg}, the best one could do is the following:
in the denominators of the prefactors, replace $m_K$ with the mass of the corresponding KK particle in the $t$-channel and $u$-channel propagator,
while in the remaining expressions use the average mass of the final state particles in place of $m_K$. 
Having made these corrections in {\sc Pythia6}, we obtained the cross-sections listed in the second-to-last column 
of Table~\ref{table:pythia}. Despite the {\it ad hoc} procedure used, the results are pretty close to the full answer from
{\sc CalcHEP}. There are two notable exceptions, the processes 313a and 315. We have checked that again the 
discrepancies are due to typos inherited from Ref.~\cite{azuelos} and left uncorrected.

\begin{table}[t]
\centering
  \begin{tabular}{ ||c||c||c|c||c|| }
  \hline
   & & {\sc Pythia6} & {\sc Pythia8} & {\sc CalcHEP}  \\  \cline{3-5}
     ISUB & Process  & \multicolumn{2}{c||}{$R^{-1}=1$ TeV} & matched spectrum  \\ \hline \hline
   311 & $g~g \rightarrow g_1^\star ~g_1^\star$   &     161.69	&     161.70	& 161.96    \\ \hline
   312 & $q~g \rightarrow {q_1^{\bullet}}~g_1^\star+c.c.$ 	  &   644.59	&     644.90	& 645.79   \\ \hline
   313 & $q~g \rightarrow {q_1^{\circ}}~g_1^\star+c.c.$    &    675.13 & 675.9	& 	676.89  \\ \hline
   314 & $q^i~q^j \rightarrow {q_1^\bullet}^i ~ {q_1^\bullet}^j + c.c.$     &   285.90 &   285.30 & 285.84	 \\ \hline
   315 & $q^i~q^j \rightarrow {q_1^\circ}^i ~ {q_1^\circ}^j + c.c.$   &    303.28	&     303.80	&  303.71  \\  \hline
   316 & $g ~g \rightarrow {q_1^{\bullet}} ~ {\bar{q}_1^{\bullet}} $   &   14.69	&    14.75	& 14.71	 \\  \hline
   317 & $g ~g \rightarrow {q_1^{\circ}} ~ {\bar{q}_1^{\circ}} $   &  17.34 & 17.37   & 17.36 \\  \hline
   318 & $q^{i} ~ {\bar{q}}^{j} \rightarrow {q_1^{\bullet i}} ~ {\bar{q}_1^{\bullet j}} $   	&     68.93	&  68.82 &   68.93 \\ \hline
   319 & $q^{i} ~ {\bar{q}}^{j} \rightarrow {q_1^{\circ i}} ~ {\bar{q}_1^{\circ j}} $   &     75.01  &   75.23	&	75.05	 \\ \hline
   320 & $q^i ~ {\bar{q}}^{j} \rightarrow {q_1^{\bullet i}} ~ {\bar{q}_1^{\circ j}} $   &  163.98 &  164.5 & 164.41   \\ \hline
   321 & $q^i ~ {q}^{j} \rightarrow {q_1^{\bullet i}} ~ {q_1^{\circ j}}  + c.c.$      & 473.71	&  474.3 & 474.43   \\ \hline
   322 &  $q ~ \bar{q} \rightarrow {q_1^{\bullet \prime}} ~ {\bar{q}_1^{\bullet \prime}} $     & 35.24 & 35.29 & 35.32  \\ \hline
   323 &  $q ~ \bar{q} \rightarrow {q_1^{\circ \prime}} ~ {\bar{q}_1^{\circ \prime}} $  &  39.91 &   40.20		& 40.20 \\ \hline
   324 & $q ~ \bar{q} \rightarrow g_1 ~ g_1 $  & 20.28 & 20.40& 20.39	 \\   
   \hline
    \end{tabular}
 \caption{ \label{tab:py68} Comparison of the strong production cross-sections (in fb) for different subprocesses in MUED
 at the 14 TeV LHC. Results are shown for our MUED implementations in 
 {\sc Pythia6} and  {\sc Pythia8}, as well as for {\sc CalcHEP} with an identical KK mass spectrum. 
 In all cases, we choose the {\tt CTEQ5L} parton distribution functions from the {\sc LHAPDF v5.9.1} library \cite{Whalley:2005nh}, 
 with $Q^2=\hat{s}$, and use the default mass spectrum as calculated in {\sc Pythia6} with $R^{-1}=1$ TeV and $\Lambda R=20$.
  } 
\end{table}

Having identified the problems with the current implementation of MUED in {\sc Pythia6}, we modified the {\sc Pythia6} 
fortran code as explained in Appendix~\ref{sec:pythia6}. In addition to fixing the typos, we generalized the treatment of the 
KK quarks, allowing for the KK doublets $q_1^\bullet$ and singlets $q_1^\circ$ to be handled independently.
Since the currently supported {\sc Pythia} distribution is the C++ {\sc Pythia8}, we also provide an implementation 
of MUED in {\sc Pythia8} which is described in Appendix~\ref{sec:pythia8}. Table \ref{tab:py68} provides a numerical cross-check 
that the results obtained with CalcHEP \cite{Datta:2010us} and with our two new versions of {\sc Pythia6} and {\sc Pythia8} are in agreement.
Notice that since we now differentiate between KK doublets $q_1^\bullet$ and KK singlets $q_1^\circ$, the number of
subprocesses in Table~\ref{tab:py68} is larger than what we had previously in Table~\ref{table:pythia}. For example,
the generic $q^\ast_1g_1^\star$ associated production {\tt ISUB=312} from Table~\ref{table:pythia} is now divided into
the production of KK doublet quarks $q_1^\bullet g_1^\star$ ({\tt ISUB=312}) and KK singlet quarks $q_1^\circ g_1^\star$ ({\tt ISUB=313})
in Table~\ref{tab:py68}.

\subsection{The relative importance of virtual level 2 KK particles}
\label{sec:level2}

The results (\ref{ggg1g1}-\ref{qqbarg1g1}) were derived ignoring diagrams containing $s$-channel propagators with level 2 KK particles,
since those always involve KK-parity conserving, but KK-number violating couplings between two SM particles and a level 2 KK particle.
Such couplings are suppressed, since they are generated at one loop, and thus one might expect the corresponding contributions to be relatively small.

\begin{figure}[t]
\includegraphics[width=.5\textwidth]{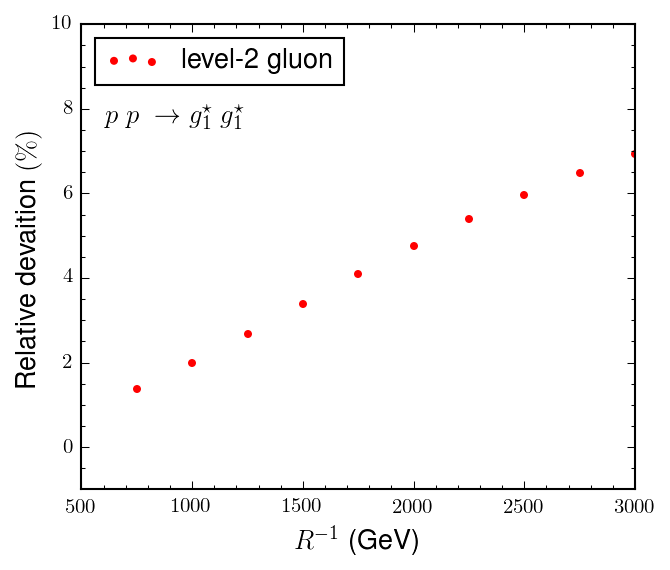}
\includegraphics[width=0.5\textwidth]{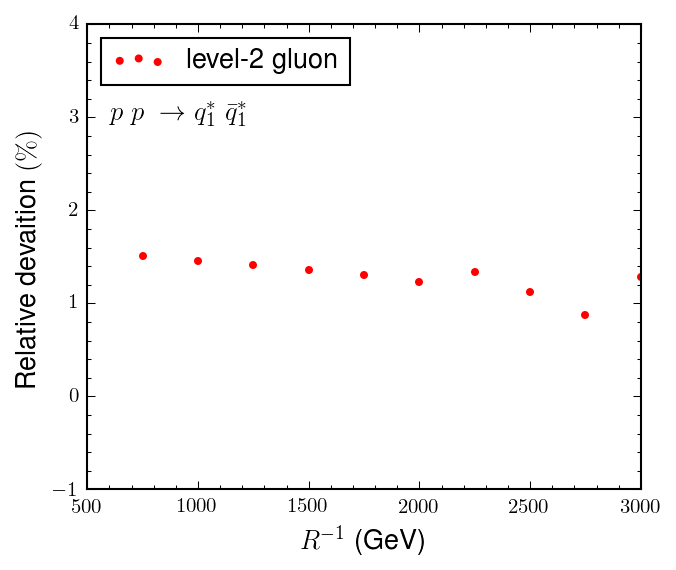}
\caption{\label{fig:ued} Relative deviation (in percent) of the cross-section for KK gluon pair-production (left) 
and KK quark pair-production, summed over all flavors, (right) after including $s$-channel propagators with virtual 
level 2 gluons. Results are plotted as a function of $R^{-1}$, for $\Lambda R=20$, $E_{cm}=14$ TeV, and choosing the {\tt CTEQ5L} PDF set with $Q^2=\hat s$.} 
\end{figure}

Fig.~\ref{fig:ued} illustrates the relative importance of diagrams with virtual level 2 KK gluons, for the case of
KK gluon pair-production (left panel) and KK quark pair-production, summed over all flavors, (right panel). 
The results, obtained with {\sc CalcHEP}, are plotted as a function of $R^{-1}$, for $\Lambda R=20$, center-of-mass energy $E_{cm}=14$ TeV, 
and choosing the {\tt CTEQ5L} PDF set. We see that the effect is at the order of a few percent, in accordance with expectations. 
 
\section{Constraints on MUED}
\label{sec:bounds}

\subsection{Constraints from cosmology}

Due to the conservation of KK parity, the LKP is stable and could be a dark matter candidate, as long as it is not charged or colored.
It would then inherit all the attractive features of a generic WIMP, and can be probed both at colliders and in dark matter experiments \cite{Cheng:2002ej}.
MUED is a very restricted model, with only two parameters: $R^{-1}$ and $\Lambda$. The relic density of the LKP depends mostly on the
mass scale of the LKP, $R^{-1}$, and less on $\Lambda$, which enters only logarithmically. Therefore, the requirement for the correct 
dark matter relic abundance singles out a preferred range for $R^{-1}$, setting a well-motivated target for the experimental searches. 
The close mass degeneracy of the level 1 KK partners, however, complicates the thermal freeze-out calculation, since one has to account for
coannihilations. The very first calculation of the MUED relic density \cite{Servant:2002aq} considered coannihilations with the level 1 KK-leptons, 
which are closest in mass to the LKP. Subsequently, the full set of coannihilation processes were also included
\cite{Burnell:2005hm,Kong:2005hn}, and the preferred mass range for the LKP (and therefore, for $R^{-1}$) 
was found to be on the order of $500-600$ GeV. However, these calculations did not include contributions from
diagrams with virtual level-2 KK particles, see Section~\ref{sec:level2}. Although the direct couplings of two SM particles to a level-2 
KK partner are loop suppressed, there is an $s$-channel resonant enhancement, and the limit on $R^{-1}$ is raised to over 1 TeV
\cite{Kakizaki:2005en,Kakizaki:2005uy}. The latest state of the art calculation of the relic density in UED was done in 
Ref.~\cite{Belanger:2010yx}, considering the KK photon $\gamma_1$ as LKP (and not the hypercharge gauge boson $B_1$ as in MUED).
The preferred range was found to be in the neighborhood of $R^{-1}\sim 1.3-1.5$ TeV, depending on the exact value of $\Lambda R$ 
and accounting for the astrophysics uncertainties.

\begin{figure}[t]
\centering
\includegraphics[width=10cm]{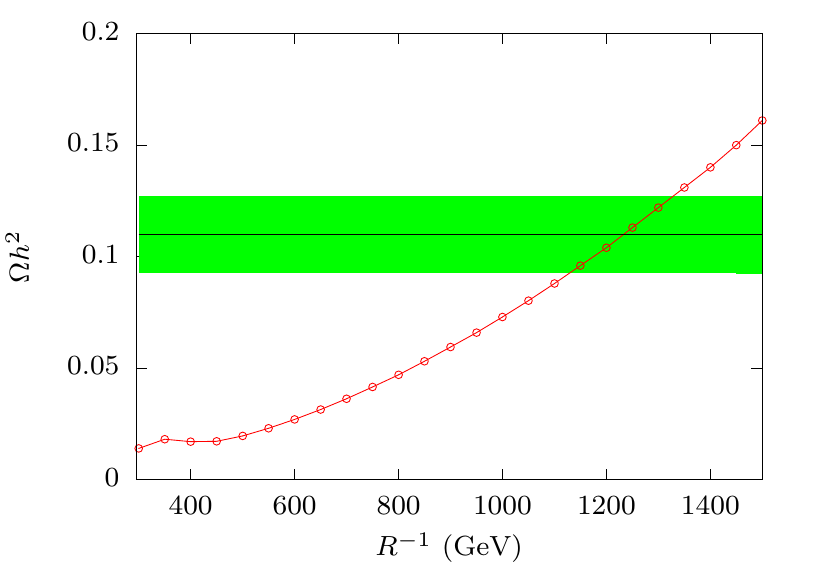}
\caption{\label{fig:relic}  The relic density in MUED as a function of $R^{-1}$, for $\Lambda R =10$, as calculated with 
{\sc micrOMEGAs} \cite{Belanger:2014hqa}, using the the MUED model files from {\sc CalcHEP} \cite{Datta:2010us}.
The horizontal green band indicates the $3\sigma$ experimentally preferred range \cite{Ade:2015xua}.
}
\end{figure}

The cosmological constraint on MUED is illustrated in Fig.~\ref{fig:relic}, where we plot the $B_1$ relic density as a function of $R^{-1}$,
for $\Lambda R =10$. We have used {\sc micrOMEGAs} \cite{Belanger:2014hqa} with the MUED model files from the 
{\sc CalcHEP} implementation \cite{Datta:2010us}. We see from Fig.~\ref{fig:relic} that in MUED, 
the correct amount of dark matter is obtained for $R^{-1}\sim 1250$ GeV. One might expect the true limit to be slightly higher, since
the plot does not include contributions from level 2 Higgs $s$-channel resonances, which are not present in the model files from 
Ref.~\cite{Datta:2010us}. The bound on $R^{-1}$ can slightly be raised further, if we allow for higher values for $\Lambda R$ than the one used in the plot.
However, in order to go above $R^{-1}\sim 1.5$ TeV would probably require some modifications to the model beyond the minimal
scenario considered here \cite{Ishigure:2016kxp}. 

\subsection{LHC simulation details}

\begin{figure}[t]
\centering
\includegraphics[width=13cm]{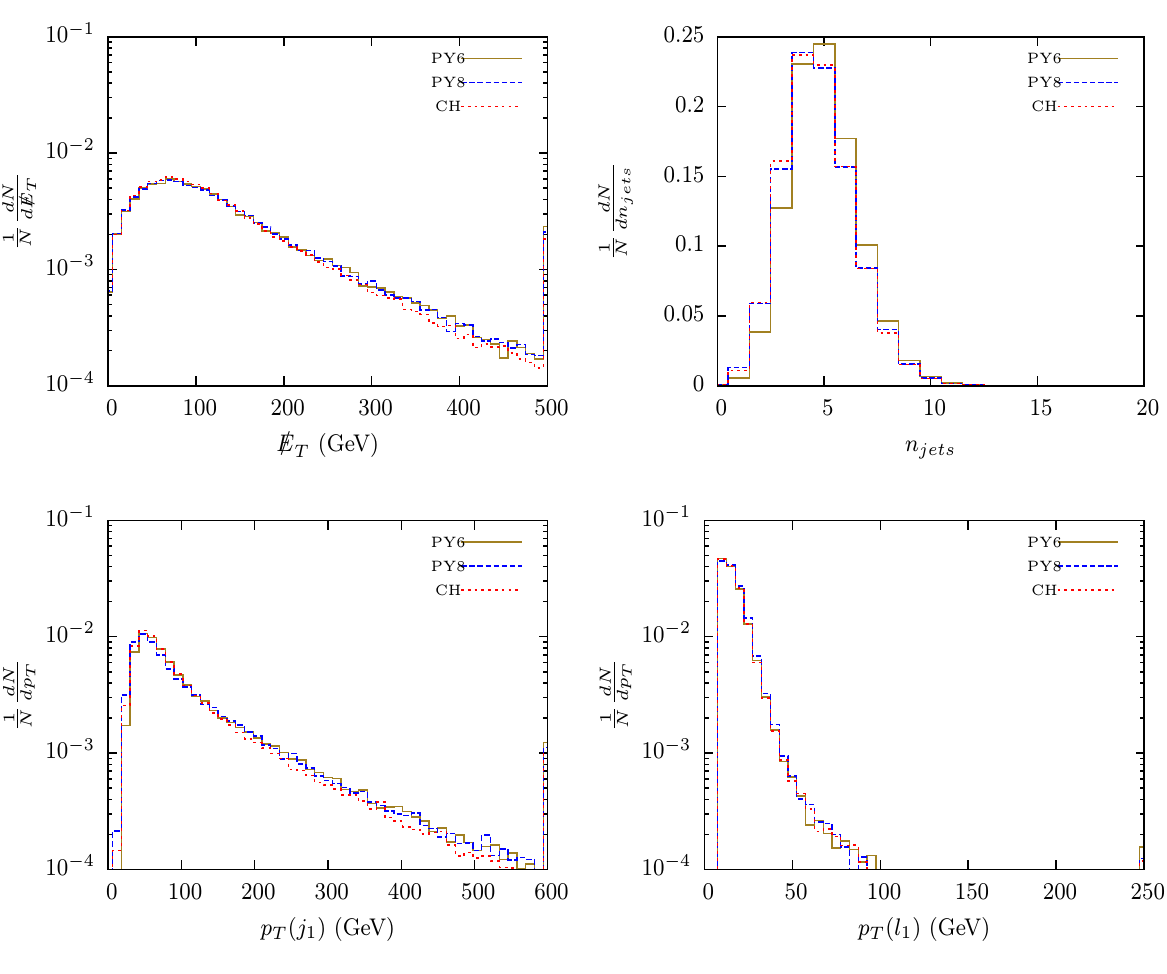}
\caption{\label{fig:kinematicsGG}  Some relevant kinematic distributions at the detector level for $g_1^\star g_1^\star$ production:
the missing transverse energy $\met$ (upper left panel), the jet multiplicity $n_{jet}$ (upper right panel),
the transverse momentum $p_T(j_1)$ of the leading jet (lower left) and
the transverse momentum $p_T(\ell_1)$ of the leading lepton (lower right).
Parton level events were produced at LHC 8 TeV with our modification of the {\sc Pythia6} code described in Appendix~\ref{sec:pythia6} (bisque solid lines),
with our {\sc Pythia8} implementation described in Appendix~\ref{sec:pythia8} (blue, long dashed lines), or 
with the MUED implementation in {\sc CalcHEP} \cite{Datta:2010us} (red, short dashed lines).
}
\end{figure}

\begin{figure}[t]
\centering
\includegraphics[width=13cm]{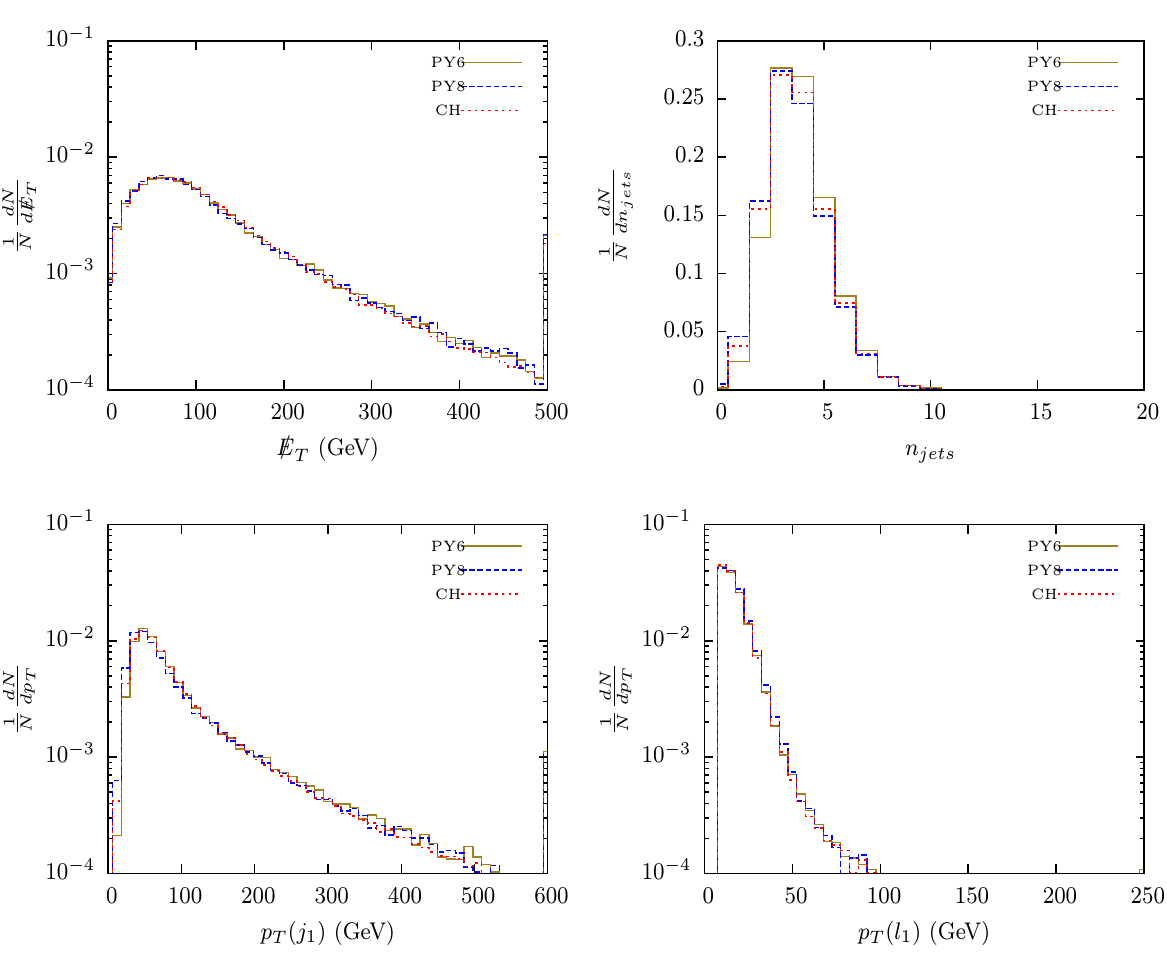}
\caption{\label{fig:kinematicsQG}  The same as Fig.~\ref{fig:kinematicsGG}, but for $g_1^\star q_1^\ast$ production.}
\end{figure}

At first glance, collider searches for MUED may appear challenging. The small mass splittings among the level 1 KK partners imply
relatively soft decay products. As in any model with a dark matter candidate, the generic signature is missing energy, $\met$ \cite{Hubisz:2008gg}.
However, $\met$ is actually measured from the total transverse momentum recoil of the visible particles in the event, and if they are relatively 
soft, the $\met$ also tends to be rather small, in spite of the large amount of {\em missing mass}.
This is illustrated in the upper left panels of Figs.~\ref{fig:kinematicsGG} and \ref{fig:kinematicsQG},
where we show the $\met$ distribution in $g_1^\star g_1^\star$ production and $g_1^\star q_1^\ast$ production, respectively.
Parton-level events were generated with our MUED implementations in {\sc Pythia6} and {\sc Pythia8} or with {\sc CalcHEP} \cite{Datta:2010us},
and run through the {\sc Pythia} event generator and the {\sc Delphes} detector simulator \cite{deFavereau:2013fsa}
as part of the standard {\sc Checkmate} simulation chain.
The figures show that the $\met$ distributions peak below 100 GeV, thus we need to consider the rest of the event in order to bring the SM backgrounds under control.

The level 1 mass degeneracy also implies that the strong production cross-sections will dominate over those for electroweak production
(see Fig.~\ref{fig:Strong-weak} for an illustrative example). The decays of KK quarks and KK gluons necessarily involve jets, 
thus we expect to have a certain amount of jets present in our signal. The upper right panels of Figs.~\ref{fig:kinematicsGG} and \ref{fig:kinematicsQG}
show histograms of the jet multiplicity $n_{jets}$, which depends on the production process: KK gluon events tend to have slightly more jets 
than KK quark events. However, those jets are not very hard, as seen in the lower left plots of Figs.~\ref{fig:kinematicsGG} and \ref{fig:kinematicsQG},
which depict distributions of the transverse momentum $p_T(j_1)$ of the leading jet in the event. Thus, while a simple multijet plus $\met$ 
search would have some reach \cite{Murayama:2011hj,Datta:2011vg,Cacciapaglia:2013wha}, 
it may be beneficial to demand in addition one or more prompt leptons from the decays of the electroweak level 1 KK bosons \cite{Cheng:2002ab}.
The lower right panels in Figs.~\ref{fig:kinematicsGG} and \ref{fig:kinematicsQG} show the corresponding distributions of the 
transverse momentum $p_T(\ell_1)$ of the leading lepton in the event. As expected, the leptons are relatively soft --- the distributions
peak at the value of the lepton $p_T$ cut used for reconstruction. Nevertheless, there is a non-negligible tail extending to high $p_T$
which opens the door for a mixed strategy targeting both jets and leptons in the final state.

The main lesson from Figs.~\ref{fig:kinematicsGG} and \ref{fig:kinematicsQG} is that when looking for MUED, there is no ``magic" cut which would 
allow an easy separation of signal from background. Yet there are several potentially useful handles which can be utilized, and the optimal
combination of jet, lepton and $\met$ requirements would be a function of the parameter space, i.e., $R^{-1}$.

\subsection{LHC bounds}

{\sc Checkmate} \cite{Drees:2013wra} is one of several tools on the market which allow for an easy recasting of published LHC data.
By now, a large number of LHC analyses from Run 1 have been implemented and validated in {\sc Checkmate}
 \cite{Aad:2013wta,Aad:2013ija,Aad:2014nua,Aad:2014qaa,Aad:2014mha,Aad:2014vma,Aad:2014pda,Aad:2014wea,Aad:2014kra,%
Aad:2014nra,Aad:2014tda,Aad:2015zva,Aad:2015wqa,Aad:2015pfx,ATLAS:2012tna,Aad:2016ett,Aad:2014bva,%
ATLAS:2013vla,ATLAS:2013qla,Aad:2014lra,TheATLAScollaboration:2013uha,Aad:2015mia,Aad:2015tin,Aad:2016wpd,Aad:2014mfk,ATLAS:2015yda,Chatrchyan:2013oev,%
Chatrchyan:2013mys,Khachatryan:2014qwa,Khachatryan:2014rra,Khachatryan:2015lwa,Khachatryan:2015nua,CMS:2013qea,CMS:2014jfa} 
and we shall make use of them in deriving the limits on MUED.\footnote{The work on incorporating Run 2 analyses in {\sc Checkmate} is underway, 
but not all of them have been implemented and validated yet.} Typically, each analysis has several search regions, with different
selections and cuts. For each signal region, {\sc Checkmate} computes the expected number of signal events $S$ after cuts, and
compares it to the $95\%$ CL upper limit $S^{95}_{exp}$ given a signal error $\Delta S$ \cite{Dercks:2016npn}. The model point is ruled out if the ratio
\beq
r \equiv \frac{S-1.96 \Delta S}{S^{95}_{exp}}
\label{rdef}
\eeq
is greater than one. 
\begin{figure}[t]
\centering
\includegraphics[width=10cm]{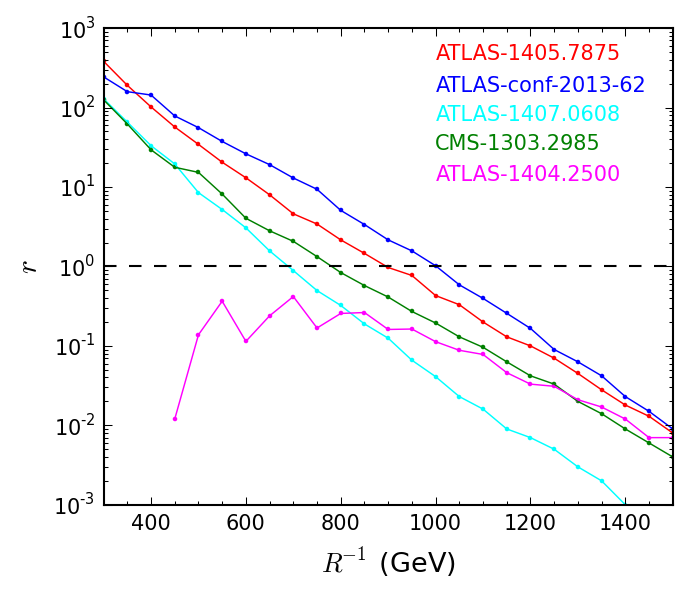}
\caption{\label{fig:lhc} The $r$-value defined in eq.~(\ref{rdef}) as a function of $R^{-1}$ (for $\Lambda R=10$), for a few representative 
analyses from {\sc Checkmate}.}
\end{figure}
In Fig.~\ref{fig:lhc} we show the $r$ values as a function of $R^{-1}$ for a few of the most sensitive LHC analyses.
The best limit (blue line in Fig.~\ref{fig:lhc}) comes from the {\tt SoftLepJ5} analysis from Ref.~\cite{TheATLAScollaboration:2013uha}, 
which requires $n_{jets} \geq 5$ with $p_T(j_1)> 180$ GeV, one lepton with $p_T(\ell_1) <25$ GeV, 
a minimum value for the magnitude of the missing transverse momentum $\mpt$ of 300 GeV, 
$m_T (\vec{p}_T(\ell),/\!\!\!\! \vec{P}_T)> 100$ GeV, and $\met>0.3\, m_{eff}$, where 
$m_{eff}$ is the scalar sum of the $\met$, transverse momenta of the jets and transverse momenta of the leptons.
Another sensitive analysis is the {\tt SR4jm} channel \cite{Aad:2014wea}
(red line in Fig.~\ref{fig:lhc}), which is an analysis with $n_{jets} \geq 4$ with
$p_T(j) > 60$ GeV, zero lepton, $\met > 160$ GeV, $m_{eff} > 1300$ GeV and $\met > 0.4\, m_{eff}$.

\section{Conclusions}
\label{sec:conclusions}

The MUED model considered here provides an interesting and motivated alternative to supersymmetry,
with unique and challenging collider phenomenology. Our two main goals in this paper were:
\begin{itemize}
\item Validation of the existing MUED implementation in the fortran version of {\sc Pythia6} and 
creating a modern implementation of MUED in the C++ version {\sc Pythia8}.
\item Demonstrating the utility of these codes when used in conjunction with a recast package like
{\sc Checkmate} in extracting LHC limits on the parameter space of the model.
\end{itemize}

Using the Run 1 LHC analyses incorporated in {\sc Checkmate}, we have derived a lower bound on $R^{-1}$ of
1 TeV from the {\tt SoftLepJ5} analysis from Ref.~\cite{TheATLAScollaboration:2013uha}. A more restrictive limit can be obtained 
from the 13 TeV analyses which are currently being implemented and validated in {\sc Checkmate}. 
Preliminary results from Ref.~\cite{DFK} indicate that the sensitivity will increase up to 1.4 TeV, 
quickly closing the window on the cosmologically motivated section of the MUED parameter space.
In addition to the pair production of level 1 KK particles, one could also use resonance searches for level 2 KK modes
to place competitive limits on the MUED parameters
\cite{Datta:2005zs,Chang:2011vt,Kim:2011tq,Chang:2012wp,Edelhauser:2013lia,Datta:2013lja,Shaw:2014gba,Dey:2014ana}.

{\bf Note added.} While we were finishing this paper, we became aware of a concurrent study,
Ref.~\cite{DFK}, whose goal was to derive bounds on the MUED parameter space from LHC data at 8 and 13 TeV,
using {\sc Herwig++} \cite{Bahr:2007ni} for event generation and {\sc Checkmate} for setting the limit.
We thank the authors of Ref.~\cite{DFK} for making a preliminary draft of their paper available to us.
In the overlapping regions of the two studies, the results were found to be in agreement.

\acknowledgments
This work is supported in part by a US Department of Energy grant DE-SC0010296. AD gratefully acknowledges generous
financial support from the Physics Department of University of Florida towards a stay on sabbatical during which this 
work was undertaken. DD acknowledges support from the University of Florida Informatics Institute in the form of a Graduate Student Fellowship.
The authors like to thank T.~Flacke, J.~S.~Kim, K.~Kong and J.~D.~Lykken for discussions. 
AD likes to thank P. de Aquino, D. Bourilkov, O. Mattelaer and A. Pukhov for helpful discussions on issues in 
handling several different packages used in this work. JB and DD thank N.~Desai, S.~Mrenna and J.~Tattersall for technical help and advice.
Participation of G.~Sarangi during the initial phase of this work is also thankfully acknowledged.

\appendix

\vskip 30pt
\section{KK gluon and KK quark production: the mass-nondegenerate case}
\label{app:matsq-ndg}
\allowdisplaybreaks
\renewcommand{\theequation}{A.\arabic{equation}}
\setcounter{equation}{0}
%
\vskip 10pt

In this appendix we present the full parton-level matrix elements for all pure QCD\footnote{We do not include diagrams mediated by 
electroweak gauge bosons and their KK modes. We omit diagrams mediated by level 2 or higher KK particles, e.g., $s$-channel 
diagrams with level two KK particles in the propagator.} strong production processes with 2 KK particles in the final state.
Here we avoid the degenerate limit approximation (\ref{degenerate}) and retain the complete dependence on all relevant mass parameters 
(listed in the third column of Table~\ref{table:pythia}). The results below have been checked in two different ways:
\begin{itemize}
\item In the degenerate limit (\ref{degenerate}) one recovers the corresponding expressions (\ref{ggg1g1}-\ref{qqbarg1g1})
given in Section~\ref{sec:degenerate}.
\item When the matrix elements were implemented in {\sc Pythia}, the corresponding cross-sections were found to agree with the numerical results 
from {\sc CalcHEP}, as shown in Table~\ref{tab:py68}.
\end {itemize}

\noindent
{\bf Process 311:} $g g \to g_1^\star g_1^\star$
\bea
\overline{\sum} |{\cal M}|^{2}_{g g \to g_1^\star g_1^\star}  &=&
\frac{9 \alpha_s^2}{16(s {\tp} {\up})^2} \left(s^2+{\tp}^2+{\up}^2\right) 
\nonumber \\
&& \quad \times  \left[6 m_{g_1^\star}^4 s^2-6m_{g_1^\star}^2 s{\tp} {\up}+3{\tp}^2 {\up}^2 + 2s^2\left({\tp}^2+{\up}^2\right)\right]
\label{Mnd311}
\eea
where ${\tp}=t-m_{g_1^\star}^2$ and ${\up}=u-m_{g_1^\star}^2$. The answer is symmetric with respect to $t'\leftrightarrow u'$.
\\
\\
{\bf Process 312:} $q g \to  {\qkb} g_1^\star$
\bea
&\overline{\sum}& |{\cal M}|^2_{q g \to {\qkb} g_1^\star}  =\dfrac{-\alpha_s^2}{72 s {\tp} {\up}^2} \left(9s^2+9{\tp}^2-{\up}^2+2\left(\amgl^2-\amqk^2\right)\left\{-s+8 \tp +8\left(\amgl^2-\amqk^2 \right )\right\}\right) \nonumber\\
&\times&  \left[2s^2+2{\tp}^2 + \dfrac{\amqk^2}{\amgl^2}{\up}^2 + \left(\amgl^2-\amqk^2\right)\left\{\amgl^2\left(-4s+2\tp+\frac{2s^2}{\tp}\right)+\amqk^2\left(2s+\tp+\frac{s^2}{\tp} \right)\right\}\right] \nonumber \\
&&
\eea
where ${\tp} = t-\amgl^2$ and ${\up}=u-\amqk^2$.
\\
\\
\\
{\bf Process 313a:} $q q \to  \qkb \qkb$
\bea
\overline{\sum}  |{\cal M}|^2_{q q \to \qkb \qkb}  &=&\frac{\alphas^2}{54 (\tp +\amqk^2-\amgl^2)^2 (\up +\amqk^2-\amgl^2)^2} \nonumber \\
&\times&
\left[ {\tp}^4 + {\up}^4 - s^4 + 6s^2 \left( {\tp}^2 + {\up}^2 \right) + \amqk^2 \left( 6 {\tp}^3 + 6 {\up}^3 -s {\tp} {\up} \right) \right.\nonumber  \\
&&+ \frac{  \amgl^2 -\amqk^2 }{\amgl^4} \left\{8 \amgl^6 \left(s^2-2s\amqk^2 \right)+8 \amgl^4 \left(s^3-3s^2 \amqk^2 +3s\amqk^4 \right) \right. \nonumber\\
 &&-  \amgl^2 \left(2{\tp}^2 {\up}^2-6\left( {\tp}^3+{\up}^3\right)+3s \tp \up \amqk^2+ \amqk^4 \left( 4 \tp \up -\frac{19}{2} s^2\right)+7s\amqk^6 \right) \nonumber \\
 &&  - 2 {\tp}^2 {\up}^2 \amqk^2 + \left.  \left. s \tp \up \amqk^4+ \amqk^6 \left(4 \tp \up -\frac{s^2}{2} \right) -s \amqk^8 \right\} \right]  
\eea
where ${\tp}=t-\amqk^2$ , ${\up}=u-\amqk^2$.
\\
\\ 
{\bf Process 313b:} $q {\qp} \to {\qkb} {\qkpb}$
\bea
\overline{\sum} |{\cal M}|^2_{q {\qp} \to {\qkb} {\qkpb}}  = \dfrac{\alpha_s^2}{18({\tp}+\amqk^2-\amgl^2)^2}\left[\dfrac{\amqk^4}{\amgl^4} {\tp}^2 +4 \amqk^2 s\left(\dfrac{\amqk^2}{\amgl^2}-2\right)+4s^2\right]
\eea
where $\tp=t-\amqk^2$.
\\
\\
{\bf Process 314:} $g g \to {\qkb} {\qbkb}$. 
\bea
\overline{\sum} |{\cal M}|^2_{g g \to {\qkb} {\qbkb}}  = 
\frac{-\alphas^2}{24 s^2 {\tp}^2 {\up}^2} 
\left(4 s^2 - 9 {\tp} {\up}\right) \left[4 m_{q_1^\ast}^4 s^2 - 4 m_{q_1^\ast}^2 s {\tp} {\up} - {\up} {\tp} \left({\tp}^2 + {\up}^2\right)\right]
\label{Mnd314}
\eea
where $\tp=t-\amqk^2$, $\up=u-\amqk^2$. 
\\
\\
{\bf Process 315:} $q \qbar \to \qkb \qbkb$
\bea
&\overline{\sum}& |{\cal M}|^2_{q \qbar \to \qkb \qbkb} =
\frac{\alphas^2}{54 s^2 \left({\tp}+{\amqk}^2-{\amgl}^2 \right)^2}
\left[12 s^4+16 s^3 \tp+23 s^2 {\tp}^2 +36 s {\tp}^3 +48 {\tp}^4 \right. \nonumber \\
&+& 12 \dfrac{\amqk^4}{\amgl^2} s \left(s^2-s \tp+4 {\tp}^2\right)+4\left(\amgl^2-\amqk^2\right)^2 \left(5s^2+12 s \tp+12 {\tp}^2+12 \amqk^2 s\right) \nonumber \\
&+& 4\left(\amgl^2-\amqk^2\right) \left(2s^3-8s^2 \tp -22s {\tp}^2-24 {\tp}^3+ \amgl^2 s^2 +\amqk^2  s\left(s-24 \tp \right)\right.\nonumber \\
&+& \left. \frac{\amgl^2-\amqk^2}{\amgl^2} \left\{s {\tp}^2 (3s-4\tp)+ \amqk^2 s \tp \left(8s -52 \tp \right)-4\amqk^4 s^2 + 3s^2 {\tp}^2 \frac{\amqk^2}{\amgl^2} \right\} \right]
\eea
where $\tp=t-\amqk^2$.
\\
\\
{\bf Process 316:} $q {\qbarp} \to {\qkd} {\qbkps}$
\bea
\overline{\sum} |{\cal{M}}|^2_{q {\qbarp} \to {\qkd} {\qbkps}} &=&
\dfrac{\alpha_s^2}{18\left(\amgl^2-{\td}-\amqd^2\right)\left(\amgl^2-{\ts}-\amqs^2 \right)} \nonumber\\
&\times & \left[4s^2+\frac{\amqd^2 \amqs^2}{\amgl^4}{\td}{\ts}-4s{\amqd}^2-4s\left\{\dfrac{\amqs^2}{\amgl^2}\left(\amgl^2-\amqd^2\right)\right\}\right]
\eea
where ${\td}=t-\amqd^2,{\ts}=t-\amqs^2$.
\\
\\
{\bf Process 317:}$q {\qbarp} \to {\qkb} {\qbkpb}$
\bea
\overline{\sum} |{\cal M}|^2_{q {\qbarp} \to {\qkb} {\qbkpb}}  = 
 \frac{{\alphas}^2}{18 \left(\amgl^2-\tp-\amqk^2\right)^2} \left[4 \dfrac{\amqk^4}{\amgl^2} s+4 s^2+8 s {\tp}+\left(4+\dfrac{\amqk^4}{\amgl^4}\right) {\tp}^2\right]
\eea
where $\tp=t-\amqk^2$. 
\\
\\
{\bf Process 318a:} $q q \to \qkd \qks$
\bea
\overline{\sum} |{\cal M}|^2_{q q \to \qkd \qks} &=& 
\frac{\alphas^2}{9 {\tp}^2 {\up}^2} \left[\frac{2\amqd^2 \amqs^2}{\amgl^2} {\sp} \left({\tp}^2 + {\up}^2 \right) + 2 {\sp}^4 - 8 {\sp}^2 \tp \up + \left(4+\frac{ \amqd^2 \amqs^2}{\amgl^4}\right) {\tp}^2 {\up}^2 \right. \nonumber \\
&+& 4\left( \amqs^4-\amqd^4 \right)\left(\amqd^2-\amgl^2 \right)^2+\left(\amqd^4-{\amqs}^4 \right)\left(\amqd^2-\amgl^2 \right)\left(2 {\tp}-{\sp}\right) \nonumber \\ 
&+& \frac{\left({\tp}^2+{\up}^2\right)}{2 \amgl^4}\left\{ 4\amgl^6 \left(\amgl^2-\amqs^2 \right)+3 \amqd^2 \amqs^2 \amgl^2 \left(\amqd^2 - \amgl^2\right) \right. \nonumber \\
&+& 3 \amqd^2 \amgl^2 \left(\amqs^4-\amgl^4\right)+\left. \amqd^2 \left(\amqd^2 \amqs^4-\amgl^6 \right)\right\} +2 \left( \amqd^4-\amqs^4 \right) {\sp} {\tp}  \nonumber \\
&+&\left(2 \amgl^2-\amqd^2-\amqs^2\right) \left(6-\frac{\amqd^2 \amqs^2}{2 \amgl^4} \right) {\sp} {\tp} {\up} \nonumber \\
&-& \left. 2\left(2\amgl^2-\amqs^2-\amqd^2\right) {\sp}^3 \right]
\eea
where $\tp=t-\amgl^2$ , $\up=u-\amgl^2$ , $\sp=s+(2\amgl^2-\amqd^2-\amqs^2)$.
\\
\\
{\bf Process 318b:} $q {\qp} \to {\qkd} {\qkps}$
 \bea
 \overline{\sum} |{\cal M}|^2_{q {\qp} \to {\qkd} {\qkps}} &=& 
\frac{-{\alphas}^2}{18\left(\amgl^2-{\td}-\amqd^2\right)\left(\amgl^2-{\ts}-\amqs^2\right)} \nonumber\\ 
&\times & \left[4 s^2 + 4 s\left({\ts}+{\td}\right)+4\dfrac{\amqs^2 \amqd^2}{\amgl^2}s+{\td} {\ts} \left(4+ \dfrac{\amqd^2 \amqs^2}{\amgl^4}\right)\right]
\eea
where $\td=t-\amqd^2,\ts=t-\amqs^2$.
\\
\\
{\bf Process 319:} $q \qbar \to {\qkpb} {\qbkpb}$
\bea
\overline{\sum} |{\cal M}|^2_{q \qbar \to {\qkpb} {\qbkpb}}=
\frac{4 \alphas^2}{9 s^2} \left(2 \amqk^2 s+ {\tp}^2 + {\up}^2 \right) 
\eea
where $\tp=t-\amqk^2  , \up=u-\amqk^2$.
\\
\\
{\bf Process 320:} $q \qbar \to g_1^\star g_1^\star$ \\
The corresponding expression is too long and we do not present it here. It can be found in the code.

\section{Implementation of MUED in {\sc Pythia6}}
\label{sec:pythia6}
\allowdisplaybreaks
\renewcommand{\theequation}{B.\arabic{equation}}
\setcounter{equation}{0}

The default implementation of MUED in {\sc Pythia6} is improved mainly along the 
following lines. 
\begin{itemize}
\item It now allows for different masses for various level 1 KK excitations.
      These may appear both in the final states and in the propagators of
      $t/u$-channel contributions. Thus, implementing correct kinematics 
      and ensuring proper interferences of diagrams require generalization
      of some of the routines handling the kinematics and incorporation of
      suitable squared matrix elements.
\item It now contains a new subprocess in the form of 
      $q \bar{q} \to g_1^* g_1^*$. A couple of production modes 
      (which originally had some subprocesses clubbed together) have now been 
      split. This action becomes necessary in view of the issue discussed in 
      the previous item.
\item For some processes, corrections in the expressions for the
      existing squared matrix elements have been made.      
\end{itemize}
%
\noindent
To achieve these, some modifications are carried out in several routines
of the original {\sc Pythia6} implementation. These are outlined below.
\begin{itemize}
\item[-] The arrays like {\tt ISET}, {\tt KFPR} and {\tt PROC} are suitably 
         modified to find and work with the newly added {\tt ISUB} entries. 
         These arrays are contained in the block {\tt PYDATA}.
\item[-] Subroutines like {\tt PYRAND} and {\tt PYSCAT} are modified to set 
         the intended initial and final states appropriately.
\item[-] The array {\tt MAPPR}, defined in the subroutine {\tt PYSIGH}, 
         stores a flag to call {\tt PYXUED}. This is modified to make 
         calls to the new {\tt ISUB} entries possible.
\item[-] The subroutine {\tt PYMAXI} calculates the maximum cross section for 
         a process. This is also modified to make calls to the new {\tt ISUB} 
         entries possible.
\item[-] The subroutine {\tt PYXUED} is modified and extended suitably to 
         include all possible $2 \to 2$ strong-processes leading to a pair
         of level 1 KK gluon and/or KK quarks. 
 \end{itemize}
%
\noindent
The new driver code `{\bf mued.f}' is copied below verbatim. It is only
different from the original driver file in having the provision for some 
extra ISUB values as shown in Table~\ref{tab:py68}. 
%
%
\begin{F77}
C...All real arithmetic in double precision.
      IMPLICIT DOUBLE PRECISION(A-H, O-Z)
C...Three Pythia functions return integers, so need declaring.
      INTEGER PYK,PYCHGE,PYCOMP
C...Parameters.
      COMMON/PYPARS/MSTP(200),PARP(200),MSTI(200),PARI(200)
C...Parameters.
      COMMON/PYDAT1/MSTU(200),PARU(200),MSTJ(200),PARJ(200)
C...Particle properties + some flavour parameters.
      COMMON/PYDAT2/KCHG(500,4),PMAS(500,4),PARF(2000),VCKM(4,4)
C...Decay information.
      COMMON/PYDAT3/MDCY(500,3),MDME(8000,2),BRAT(8000),KFDP(8000,5)
C...Selection of hard scattering subprocesses.
      COMMON/PYSUBS/MSEL,MSELPD,MSUB(500),KFIN(2,-40:40),CKIN(200)
      COMMON/PYPUED/IUED(0:99),RUED(0:99)
C...EXTERNAL statement links PYDATA on most machines.
      EXTERNAL PYDATA
      DOUBLE PRECISION rinv_val
C...Events to be generated:
        nev = 50000
        rinv_val=500.0
        pmas(25,1) = 125.0d0
        pmas(6,1) = 173.1d0
C...First section: Initialize MUED.
       IUED(1)=1
       IUED(2)=0
       IUED(3)=5
       IUED(5)=1
       IUED(6)=1
C...Set MUED input parameters.
       RUED(1)=rinv_val
       RUED(4)=20.d0
C...Select generic MUED generation.
       MSEL=0
c... ISUB is running from  311-324
       MSUB(312) = 1  !  q + g -> q*_D + g*
       MSUB(313) = 1  !  q + g -> q*_S + g*
C...If interested only in cross sections and resonance decays:
C...switch off initial and final state radiation,
C...multiple interactions and hadronization.
      MSTP(61)=0
      MSTP(71)=0
      MSTP(81)=0
      MSTP(111)=0
      MSTP(32)=4  ! Q^2=s_hat
* LHAPDF settings
       MSTP(52)=2        ! Interface LHAPDF
c-----------------------------------------------------------
       MSTP(51) =19070  ! for CTEQ5L (LO fit) LHAPDF
c-----------------------------------------------------------
C...Initialization for the Tevatron or LHC.
      CALL PYINIT('CMS','p','p',14000D0)
C...Loop over the number of events.
      DO 200 IEV=1,nev
        IF(MOD(IEV,10000).EQ.0) WRITE(6,*)
     &  'Now at event number',IEV
C...Event generation.
        CALL PYEVNT
C...List first few events.
          IF(IEV.LE.50) CALL PYLIST(1)
  200 CONTINUE   ! end of event Loop
C...Cross section table.
      CALL PYSTAT(1)
      WRITE(*,'(I10,3x,10f17.10)') isub_val,PARI(1)*1.0e12
        END
\end{F77}

The necessary files for the improved MUED implementation in {\sc Pythia6}
can be downloaded from \url{http://www.hri.res.in/~jyotiranjan/download.html}.

\section{Implementation of MUED in {\sc Pythia8}}
\label{sec:pythia8}

It is easier to add new models to {\sc Pythia8} (when compared to {\sc Pythia6}) 
thanks to its highly modular structure. In our implementation of the MUED scenario, we 
broadly adhere to the strategy followed for SUSY in {\sc Pythia8}.
We retain the internal routines of {\sc Pythia8} and just add plugins for our purpose. 
{\sc Pythia8} depends on other spectrum generators to find the spectrum it likes to work with. Thus, the 
MUED inputs and the resulting spectrum (along with other related information) 
are fed to {\sc Pythia8} from {\sc CalcHEP} via {\sc SLHA} format. However, one 
can indicate and use input masses of his/her choice directly in the code. 
The {\sc SLHA} file also contains the information on the decays (widths and branching 
fractions) of all the level 1 KK excitations. 
To compare the {\sc Pythia8} results with those from {\sc CalcHEP}, we use 
$\alpha_s$  obtained directly from the {\sc LHAPDF} parton distribution in use. 
However, $\alpha_s$ from {\sc Pythia8} itself can be used just by toggling 
the switch {\tt GetAlphasFromPDF(true)} to 
{\tt GetAlphasFromPDF(false)}.

\begin{figure}[t]
\includegraphics[width=0.5\textwidth]{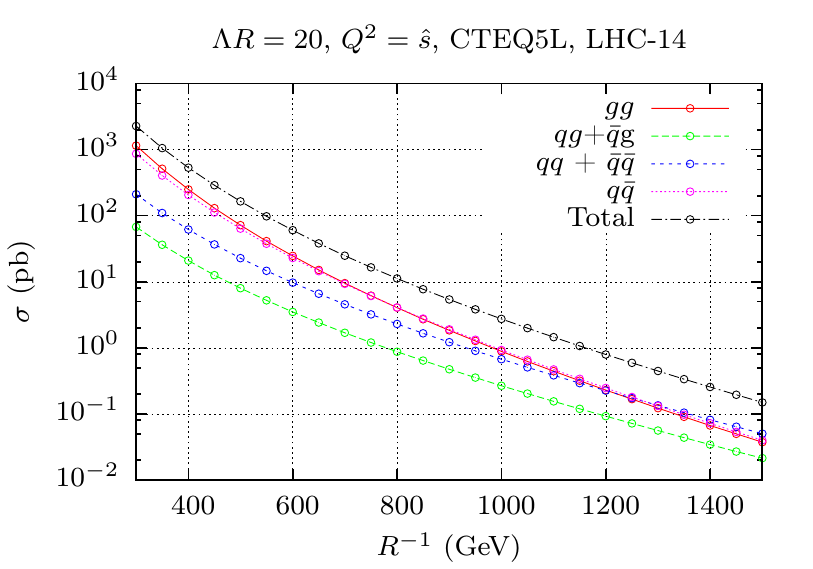}
\includegraphics[width=0.5\textwidth]{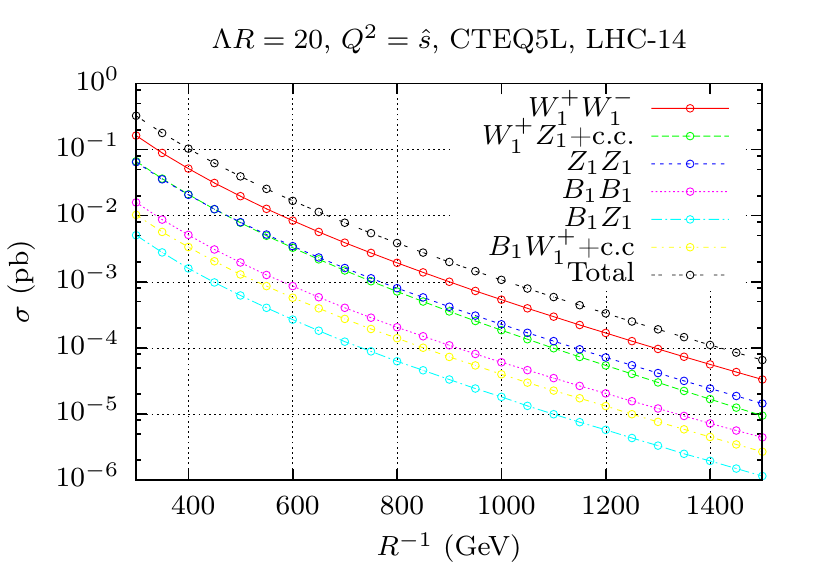}
\caption{\label{fig:Strong-weak} Left(Right): Production cross sections of strong (weak) KK particles at 14 TeV LHC
for $\Lambda R=20$ as a function of $R^{-1}$. The individual contributions are categorized by the type of initial state partons (left panel)
or final state particles (right panel). } 
\end{figure}

We have made an extensive use of {\tt ResonanceWidths} class and 
{\tt Sigma2Process} class to implement resonant 2-body decays and 
$2\rightarrow2$ hard scattering processes, respectively.
\begin{table}[t]
	\centering
	\begin{tabular}{|c|c|c|}
		\hline
		Sub Processes (including c.c.)& {\sc Pythia8} identifier  &  {\sc Pythia8} class\\  \hline
		$g g \rightarrow g_1^\star g_1^\star$ 	  & {\tt gg2KGKG}  & {\tt Sigma2gg2KGKG}  \\ \hline
		$q_i g \rightarrow q_i^\bullet  g_1^\star,q_i^\circ  g_1^\star $ &  {\tt qg2KQKG}  &  {\tt Sigma2qg2KQKG}\\ \hline
		$q_i q_j \rightarrow q_i^\bullet q_j^\bullet, q_i^\circ q_j^\circ, q_i^\bullet q_j^\circ$ &    {\tt qq2KQKQ} &    {\tt Sigma2qq2KQKQ} \\ \hline
		$q_i \bar{q}_j \rightarrow q_i^\bullet \bar{q}_j^\bullet, q_i^\circ \bar{q}_j^\circ, q_i^\bullet \bar{q}_j^\circ$&  {\tt qqbar2KQKQbar}  &  {\tt Sigma2qqbar2KQKQbar} \\ \hline
		$q_i \bar{q}_i \rightarrow q_j^\bullet \bar{q}_j^\bullet,  q_j^\circ \bar{q}_j^\circ $ ($i \neq j$)&  {\tt qqbar2KQpKQpbar} &  {\tt Sigma2qqbar2KQpKQpbar} \\ \hline
		$q \bar{q} \rightarrow g_1^\star g_1^\star$ &  {\tt qqbar2KGKG} &  {\tt Sigma2qqbar2KGKG} \\ \hline
		$g g \rightarrow q^\bullet \bar{q}^\bullet, q^\circ \bar{q}^\circ$ & {\tt gg2KQKQbar} & {\tt Sigma2gg2KQKQbar} \\ \hline
		\multirow{1}{*}{$q_i \bar{q}_j \rightarrow V_1 V_1$}&  {\tt qqbar2W1W1} &  {\tt Sigma2qqbar2W1W1}\\ \hline
		&  {\tt qqbar2Z1Z1} &  {\tt Sigma2qqbar2Z1Z1}\\ \hline
		&  {\tt qqbar2Z1W1} &  {\tt Sigma2qqbar2Z1W1}\\ \hline
		&  {\tt qqbar2B1V1} &  {\tt Sigma2qqbar2B1V1}\\ \hline
	\end{tabular}
	\caption{\label{tab:py8-procs} The generic processes implemented in {\sc Pythia8}.}
\end{table}
In {\sc Pythia8} we incorporate all the MUED processes implemented in the 
original version of {\sc Pythia6} plus the missing ones that we added in our
{\sc Pythia6} implementation. In addition, we included pair-production
processes for level 1 KK gauge bosons. For pair-production of level 1 KK quarks, 
processes involving only strong interaction are considered. Processes that involve 
virtual level-2 gauge bosons are not included as of now, as is also the case with {\sc Pythia6}. 
The generic $2 \to 2$ scattering processes implemented in {\sc Pythia8} are listed 
in Table \ref{tab:py8-procs} and the corresponding production cross-sections 
are illustrated in Fig.~\ref{fig:Strong-weak}, as a function of $R^{-1}$, for $\Lambda R =20$.
%
%

The driver file `{\bf mued.cc}' for working with the MUED scenario is pasted 
below verbatim. The example process indicated there is
$q g \rightarrow q_1^\bullet g_1^\star,q_1^\circ g_1^\star$. 
\vskip 15pt
\begin{CPP}
#include "Pythia8/Pythia.h"
#include "SqmeMUED.h"
#include <unistd.h>
using namespace Pythia8;

int main(int argc, char* argv[])
{
Pythia pythia;
int nEvent = 200000;
int nAbort = 0;

pythia.settings.flag("PartonLevel:FSR",false);
pythia.settings.flag("PartonLevel:ISR",false);
pythia.settings.flag("HadronLevel:all", false);
pythia.settings.flag("PartonLevel:MPI", false);
pythia.readString("HadronLevel:Hadronize = off");
pythia.readString("HadronLevel:Decay = off");

pythia.readString("Check:levelParticleData = 12");
pythia.readString("SigmaProcess:renormScale2 = 4");
pythia.readString("SigmaProcess:factorScale2 = 4");
pythia.readString("PDF:pSet =    LHAPDF6:CT10/0");
// pythia.readString("PDF:pSet = LHAPDF5:CT10.LHgrid");

pythia.readString("Beams:eCM = 14000");
pythia.readString("SLHA:file = decaySLHA_1000.txt");
pythia.readString("SLHA:useDecayTable = off");

//to match with SLHA pdg code
int pdgDoublet=5500000; int pdgSinglet=6500000;
Sqme2MUED mued(pdgDoublet,pdgSinglet,pythia.settings);
mued.GetProcessType("qg2KQKG");
//mued.GetProcessType("gg2KGKG"); 
mued.GetAlphasFromPDF(true) ;
for(int i=0; i< mued.GetSigmaPtr().size();i++)
pythia.setSigmaPtr(mued.GetSigmaPtr()[i]);

pythia.init();// Initialization for LHC.
int iAbort = 0;

for (int iEvent = 0; iEvent < nEvent; ++iEvent) // Begin event loop.
{
if (!pythia.next()) // Generate events. Quit if failure.
{
if (++iAbort < nAbort) continue;
cout << " Event generation aborted prematurely, owing to error!\n";
break;
}
}
pythia.stat();// Final statistics.
return 0;
}
\end{CPP}
%
\begin{table}
\begin{tabular}{|c|c|c|c|c|c|c|}
	\hline
	Particles & $g_1$ &$B_1$   & $Z_1$ &$W_1^{\pm}$&  $q_1^\bullet$ $(\ell_1^\bullet)$ & $q_1^\circ$ $(\ell_1^\circ)$ \\ \hline
    (PDG) ID & 5500021 & 5500022 & 5500023 & 5500024 & 5500000 &6500000 \\
      &  &  &  &  & +PDG($q/\ell$)&+PDG($q/\ell$) \\ \hline
\end{tabular}
\caption{\label{tab:pdg}PDG codes of the KK particles. Here PDG($q/\ell$) stands for the PDG code of a SM quark $q$ or a SM lepton $\ell$.}
\end{table}
\noindent
The PDG IDs assigned to various level 1 KK excitations are shown in Table 
\ref{tab:pdg}. 

The necessary files for the MUED implementation in {\sc Pythia8}
can be downloaded from \url{http://www.hri.res.in/~jyotiranjan/download.html}.



\end{document}